\documentclass{emulateapj}
\usepackage{graphicx,url,amssymb,amsmath,longtable,rotating,color,units,wasysym,subfigure,epsfig,listings,bm, natbib}

\slugcomment{Submitted for publication in The Astrophysical Journal}
\shorttitle{Gravitomagnetic waves from newborn neutron stars: sensitivity study}
\shortauthors{Mytidis, Coughlin and Whiting}

\newcommand{\be}{\begin{eqnarray}}
\newcommand{\ee}{\end{eqnarray}}

\begin{document}

\title{Constraining the r-mode saturation amplitude from a hypothetical detection of r-mode gravitational waves from a newborn neutron star - sensitivity study} 

\author{Antonis Mytidis\altaffilmark{1} Michael Coughlin\altaffilmark{2} and Bernard Whiting\altaffilmark{3}}

\altaffiltext{1,3}{Department of Physics, University of Florida, 2001 Museum Road, Gainesville, FL 32611-8440; mytidis@phys.ufl.edu, bernard@phys.ufl.edu}
\altaffiltext{2}{Department of Physics, Harvard University ; coughlin@physics.harvard.edu} 

\begin{abstract}

This paper consists of two related parts: In the first part we derive an expression of the moment of inertia (MOI) of a neutron star as a function of observables from a 
hypothetical r-mode gravitational wave detection. For a given r-mode detection we show how the value of the MOI of a neutron star constrains the equation of state 
(EOS) of the matter in the core of the neutron star. Subsequently, for each candidate EOS, we derive a possible value of the saturation amplitude, $\alpha$, of the 
r-mode oscillations on the neutron star. Additionally, we argue that a r-mode detection will provide clues about the cooling rate mechanism of the neutron star. 
The above physics that can be derived from a hypothetical r-mode detection constitute our motivation for the second part of the paper. In that part we present 
a detection strategy to efficiently search for r-modes in gravitational-wave data. R-mode signals were 
injected into simulated noise colored with the advanced LIGO (aLIGO) and Einstein Telescope (ET) sensitivity curves. The r-mode waveforms used are those 
predicted by early theories based on a polytropic equation of state (EOS) neutron star matter \citep{GWHYNS}. In our best case scenario ($\alpha$ of order $10^{-1}$),
the maximum detection distance when using the aLIGO sensitivity curve is $\unit[\sim1]{Mpc}$ (supernova event rate of 3-4 per century) 
while the maximum detection distance when using the ET sensitivity curve is $\unit[\sim10]{Mpc}$ (supernova event rate of 1-2 per year). Our results suggest that 
if aLIGO takes data in 2015, it may be possible to set constraints for the EOS of the neutron star remnant of the Messier 82 supernova (SN2014J) that occurred 
in January 2014. Depending on the r-mode parameters and detection sensitivity we may be able to determine an upper bound (in the worst case) on the r-mode saturation 
amplitude or (in the best case) determine its value based on the assumed EOS. \\ 

\keywords{aLIGO, neutron stars, equation of state, r-modes, gravitomagnetic waves, gravitational waves, supernova, CFS instability}

\end{abstract}

\section{Introduction}
A hypothetical detection of r-mode gravitational radiation from newborn neutron stars could have at least three major implications in our understanding of neutron stars: 
(i) explanation of the low rotational frequencies of the observed neutron stars when compared to their possible rotational frequencies at birth, 
(ii) set constraints on the equation of state of the matter in the core of the neutron star and (iii) set upper bounds on $\alpha$ and settle the debate about the 
magnitude of the saturation amplitude of the r-mode mass current oscillations on neutron stars. \\

After some preliminaries in section 2, the paper splits into two parts: the first part consists of sections 3-7 and the second part consists of sections 8-11. 
In the first part we present the physics of a neutron star that can be derived from a hypothetical r-mode detection: (i) the MOI of the neutron star, 
(ii) the EOS of the matter in the neutron star nucleus and (iii) the saturation amplitude ($\alpha$) of the r-mode mass-current oscillations. 
These results constitute our motivation for the second part, to perform a sensitivity study on r-modes: using (time-shifted) eLIGO data, recolored with aLIGO and ET 
sensitivity curves, we examine the distances at which our decision making algorithms can be sensitive to r-mode signals from newborn neutron stars with a False Alarm Rate (FAR) 
of $0.1 \%$ and a False Dismissal Rate (FDR) of $50\%$. \\

The first part of the paper is organized as follows: in section 3 we discuss the power dependence on the gravitational-wave frequency and the saturation amplitude, $\alpha$, 
of the r-mode oscillations. This result is used in section 4 to argue that among all possible r-mode sources newborn neutron stars are the most promising sources 
of detectable r-mode gravitational waves. In section 5, we present the time frame after a supernova explosion in which we would expect to detect r-mode gravitational waves.  
We end the first part of the paper with sections 6 and 7, where we present the motivation for a r-mode gravitational-wave search. In section 6 we derive a relation 
between r-mode gravitational waves and the MOI of the neutron star, while in section 7 we show how this relation is used to constrain the EOS of the neutron star 
matter and subsequently set constraints on the saturation amplitude, $\alpha$, of neutron star r-mode oscillations. \\

The second part of the paper is organized as follows: in sections 8 and 9 we discuss the choice of parameters used to construct the waveforms needed to design the sensitivity study.
In section 8 we determine the range of possible values for the initial spindown frequency, $\Omega_o$, of the neutron star,
while in section 9 we argue about the choice of the range of values for the parameter $\alpha$.
These 2 parameters determine the waveforms used in the sensitivity study presented in sections 10 and 11. In section 10 we present our choice of waveforms and
plot several of them demonstrating the frequency spin-down dependence on $\alpha$ and $f_o$, while in section 11 we present the method of waveform 
injection and recovery as well as the results we obtained on the detection distances for each waveform that was used. 
The paper ends with a discussion on the results from our sensitivity study as well as suggestions for future work that will follow in a second paper. \\

\section{Preliminaries}

In the late 1990's, the r-mode toroidal pulsations of a neutron star became very promising for generating strong gravitational-wave signals due to the 
Chandrasekhar-Friedman-Schutz (CFS) instability they exhibit \citep{LAGRANGEPERT,SECINSTROTNS}. R-modes of any harmonic, frequency and amplitude are 
subject to this instability at any angular velocity of the star \citep{NCRM1997,JFSM1998}. Therefore, even the smallest toroidal perturbations in the velocity of 
the neutron star mass currents will keep increasing in amplitude. Since the energy source of the r-mode oscillations is the rotational energy of the star, 
these small perturbations can eventually reach energy values of the order of the rotational energy of the neutron star. That would imply that all neutron 
stars are unstable and would contradict our observations. Therefore, some r-mode damping mechanism must be active resulting in 
the r-mode oscillation reaching a saturation amplitude, thus preserving the stability of the neutron star. \\

In considering the saturation amplitude, its normalization is such that values of order 1 carry energy of the same order of magnitude as the total rotational energy of the neutron star.
Some authors have introduced damping mechanisms that can cause saturation at r-mode oscillation amplitudes of order $10^{-4}-10^{-2}$ dimensionless 
units \citep{SDNS2009}, while others have introduced mechanisms that cause saturation at amplitudes equal to or larger than $10^{-1}$ \citep{ALFORD}. 
On the other hand, some studies have examined the hypothesis 
that, under certain conditions, r-mode oscillations may get suppressed when neutron star models with a solid crust are considered \citep{SCSRM}. In this case, the r-modes would 
be completely suppressed when the temperature of the neutron star drops below $\unit[10^8]{^oK}$. However, we are interested in the very early stages in the life of a newborn neutron 
star, when its core temperature is still around $ \unit[10^9-10^{10}]{^oK}$. Therefore, this solid-core suppression, if it exists, does not come into play until much later on in the evolution 
of the neutron star. The uncertainty exemplified in this situation calls for the design of a search for a r-mode gravitational radiation from newborn neutron stars to discover their saturation amplitudes.\\

The Bondarescu, Teukolsky and Wasserman '09 result was the outcome of numerous earlier studies e.g. \citep{RMCOUPLENERGTRANS} that discovered a coupling
between the r-mode and other inertial modes. Bondarescu et al.'s
numerical study showed that for initial amplitudes of order $10^{-6}$, non-linear couplings of the r-mode with other 
(neighboring in frequency) inertial modes can result in saturation of the r-mode amplitude at values of order up to $10^{-2}$ 
(in their most optimistic case scenario). \\

If the assumptions in the Bondarescu et \text{al.\,\,}model do not hold and 
the r-mode does not saturate at their suggested amplitudes, a different mechanism may saturate the r-mode oscillations at higher 
amplitudes. Such possibilities were considered by Mark Alford et al., who worked on models where the r-mode oscillations can reach saturation amplitudes 
2 to 3 orders of magnitude larger than those predicted by Bondarescu et al. They explored mechanisms that can saturate the r-mode oscillations 
at amplitudes of order $10^{-1}$ - $1$ \citep{ALFORD}. They discovered that non-linear bulk viscosity will cause r-modes to saturate at large amplitudes and spin 
down the neutron star in much shorter time scales than Bondarescu et \text{al.\,\,}had suggested. \\

\section{Energetics}
\label{sec:models}

In this section we present the expression for the r-mode waveforms that was derived in \citep{GWHYNS} specifically 
for neutron star matter obeying a polytropic EOS. We then derive the power dependence on the r-mode gravitational wave frequency, and the saturation amplitude $\alpha$. 
This is used in section 4 where we argue that newborn neutron stars are the most promising sources for r-mode gravitational waves.\\ 

\subsection{Frequency evolution and energy of the r-mode oscillation}

Though very simplistic, the \text{Owen et al.\,\,\,}'98 model is still a very good approximation for the early stages of the neutron star spin-down \citep{GWHYNS}.
More complicated numerical methods have shown that a r-mode saturation amplitude $\alpha=10^{-2}$ can result in a spin-down whose energy loss can be detected
as gravitational radiation by aLIGO \citep{SDNS2009}. When this saturation amplitude is used in the '98 model, we see that there is a good agreement in the
angular velocity evolution of the neutron star up to several months after the start of the neutron star spin-down.
The evolution of the angular velocity $\Omega$ of the neutron star in the Owen et al. '98 model is described by

\begin{equation}
 \label{eq.f}
 \frac{d \Omega}{dt} = \frac{2 \Omega}{\tau_{GR}} \frac{\sigma Q}{1-\sigma Q}
\end{equation}

\noindent
with $\sigma=\alpha^2$, where $\alpha$ is the amplitude at which the r-mode oscillation saturates, $Q$ is a dimensionless equation of state 
dependent parameter that takes the value $Q=9.4\times 10^{-2}$ for the polytropic neutron star model and $\tau_{GR}$ is the gravitational
radiation (e-folding) time scale and is given by

\begin{equation}
 \label{eq.s}
 \frac{1}{\tau_{GR}} = \frac{1}{\tilde{\tau}_{GR}} \left (  \frac{\Omega^2}{\pi G \tilde{\rho}}   \right)^{l+1}
\end{equation}

\noindent
where $G$ is the gravitational constant and $\tilde{\rho}$ is the average density of the neutron star \citep{GWHYNS}. Combining \eqref{eq.f} 
and \eqref{eq.s} for $l=2$, $\tilde{\tau}_{GR}= \unit[-3.3]{s}$ and neglecting terms with powers of $\sigma Q$ higher than 1 we get

\begin{equation}
\label{eq.t}
 \frac{d \Omega}{dt} = -\frac{2 \lambda Q}{3.3} \frac{\Omega^7}{(\pi G \tilde{\rho})^3}
\end{equation}

\noindent
where $\lambda = \sigma s^{-1}$. Integrating \eqref{eq.t} we get the angular velocity evolution of the neutron star to be

\begin{equation}
\label{eq.fo}
 \Omega(t) = \frac{\Omega_{o}}{\left ( 1 + \frac{12 \lambda Q}{3.3} \left ( \frac{\Omega_{o}^2}{\pi G \tilde{\rho}}\right )^3  t \right)^{ \frac{1}{6} }}
\end{equation}

\noindent
where $\Omega_o$ is the angular velocity at birth. An extension of this model that includes magnetic braking effects, is described 
in \citep{HOLAI} and \citep{HKTOP}. From \eqref{eq.fo}, we can derive the time evolution of the neutron star rotational frequency 
(with $f_o$ being the initial frequency) to be

\begin{equation}
\label{eq.6}
  f(t) = \frac{f_o}{ \left( 1 +  \frac{12(2\pi)^6 \lambda Q}{3.3} \left( \frac{f_o^2}{\pi G \tilde{\rho}} \right )^3 t \right)^{ \frac{1}{6} } } 
\end{equation}

\noindent
Dividing numerator and denominator of the right hand side by $f_o$ we get 

\begin{equation}
 \label{eq.6}
  f(t) = \frac{1}{ \left( \frac{1}{f_o^6} +  \frac{7.3\times10^3 \lambda Q}{(G \tilde{\rho})^3} t \right)^{ \frac{1}{6} } } 
\end{equation}

\noindent
which results in 

\begin{equation}
 f(t) = \frac{1}{ \left ( f_o^{-6} + \mu t \right )^{\frac{1}{6}} }
\end{equation}

\begin{equation}
\label{eq.7}
  \mu= 1.1 \times 10^{-20} |\alpha|^2 \frac{\text{s}^{-1}}{\text{Hz}^6}
\end{equation}

\noindent
and the time needed for the neutron star rotational frequency to evolve from some initial frequency $f_o$ to a frequency value $f < f_o$ (using $f_o^6 \gg f^6$)
is given by

\begin{equation}
\label{eq.8}
 t \simeq \frac{93}{|\alpha|^2} \left( \frac{1 \text{kHz}}{f} \right)^6 \text{s} 
\end{equation}

The Owen et al '98 model depends on two parameters: the initial angular velocity $\Omega_o=2\pi f_o$ discussed in section 2 and the r-mode oscillation
saturation amplitude $\alpha$. Using the upper bound of the angular velocity $\Omega_{ns} \unit[\simeq7.2\times10^3]{rad\,s^{-1}}$ (discussed in section 8.1) we conclude that we can take 
$f_o \unit[\simeq1.1\times10^3]{Hz}$. Since the r-mode gravitational wave frequency, $f_{gw}$, is related to the neutron star rotational frequency, $f_{ns}$, by $f_{gw}=4/3f_{ns}$ 
we conclude that r-mode gravitational waves from newborn neutron stars would have initial frequencies bounded above by $f_{gw}\unit[\simeq1.5\times10^3]{Hz}$. \\

Estimating the value of the parameter $\alpha$ is not as easy as estimating $f_o$. During the decade following the '98 paper there was no 
known mechanism that could stop the r-mode amplitude from growing due to the CFS instability. Therefore, the r-mode saturation amplitude was 
assumed to take values of order 1. \\

\begin{figure}
\epsscale{1.2}
\plotone{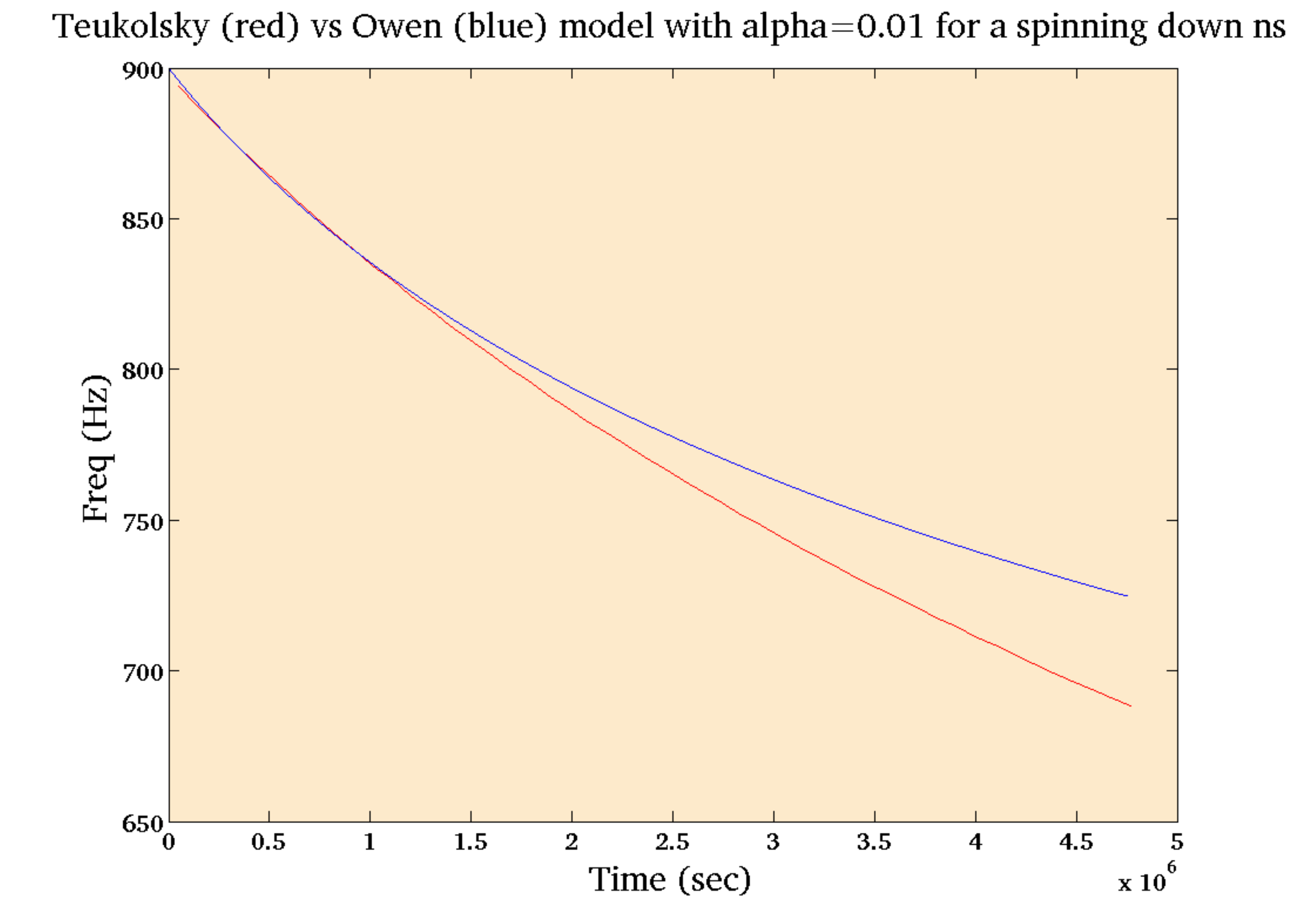}
\caption{R-mode waveform (red) predicted by the best case scenario of Bondarescu et al. numerical simulations where the r-mode oscillations saturate 
at an amplitude $\alpha$ of order $10^{-2}$. For comparison this is plotted on the same axes with the waveform described by the Owen et \text{al.\,\,}model with $\alpha=10^{-2}$.
The two waveforms are in a very close agreement during the early stages (up until $\unit[\sim10^6]{s}$ ) of the frequency evolution. This is a good indication
that even though the Owen et \text{al.\,\,}model is very simplistic and deviates from complicated numerical simulations for the total duration of the spin-down, 
it may be very accurate for the early stages of the frequency spin-down. In our sensitivity study we used $\unit[2.5\times10^3]{s}$ waveforms from the Owen et al model.} \label{Fig:ef1}
\epsscale{1.0}
\end{figure}

This is not the case in later simulations (Bondaresku, Teukolsky, Wasserman, 2009) where the r-mode oscillations exhibit non-linear couplings with 
daughter modes and the saturation amplitude ($\alpha$) reaches maximum values of order $10^{-3}$ - $10^{-2}$. The models that predict these saturation amplitudes
are discussed in section 9.1. Data from the Bondarescu et al. model have been obtained and a comparison of the two evolutions (both cases with $\alpha = 10^{-2}$) is 
given in Fig.\ref{Fig:ef1}. \\

To estimate the energy stored in a neutron star r-mode and the power radiated as r-mode gravitational waves we need to know the values of $f_{gw}$ and $\alpha$. 
From Owen et \text{al.} '98 \citep{GWHYNS} the energy of the $l=2$ r-mode (with $M=1.4$ solar masses, $R=\unit[12]{km}$) is given by 

\begin{equation}
\label{eq.11}
 E_r = 0.82 \times 10^{-2} |\alpha|^2 M R^2 \Omega^2 = 1.3 \times 10^{38} f_{ns}^2 |\alpha|^2
\end{equation} 

\noindent
and using $f_{gw}=4/3 f_{ns}$ we find 

\begin{equation}
\label{eq.12}
 E_r = 0.73 \times 10^{38} f_{gw}^2 |\alpha|^2
\end{equation}

\noindent
For $f_{ns}= \unit[1.5\times10^3]{Hz}$ and $\alpha=1$ we find that the energy of the r-mode of a newborn neutron star is of order $\unit[10^{51}]{ergs}$. \\

\subsection{Power dependence on the r-mode radiation frequency and satuarion amplitude $\alpha$}

The gravitational-wave amplitude $h_o$ of a signal with frequency $f_{gw}=f$ at a distance $d$ and power radiated $\dot E$ is given by \citep{SDNS2009}

\begin{equation}
\label{eq.13}
 h_o^2= \left( \frac{5G}{2 \pi^2 c^3} \right) \left ( \frac{1}{d^2} \right) \left( \frac{1}{f^2} \right ) \dot E
\end{equation}

\noindent
(in mks units) or 

\begin{equation}
\label{eq.14}
 h_o^2 \sim 6.3 \times 10^{-37} \left( \frac{1}{d^2} \right) \left( \frac{1}{f^2} \right ) \dot E
\end{equation}

\noindent
The gravitational-wave amplitude at a distance $d$, the gravitational-wave frequency and the r-mode oscillation amplitude $\alpha$ are related by \citep{HOLAI}

\begin{equation}
\label{eq.15}
 h_o \approx 1.5 \times 10^{-23}  \left( \frac{1\text{Mpc}}{d} \right)  \left( \frac{f}{1\text{kHz}}  \right )^3 | \alpha |
\end{equation}

\noindent
Substituting \eqref{eq.15} in \eqref{eq.14}, we get a relation between the r-mode amplitude, the frequency of gravitational radiation and the power radiated in r-mode gravitational waves

\begin{equation}
\label{eq.16}
\dot E \approx 3.5 \times 10^{19} f^8 |\alpha|^2
\end{equation}

\noindent
Since the gravitational radiation power is proportional to the $8^{th}$ power of frequency we expect a very rapid decrease of the signal power. 
From \eqref{eq.16}, we see that by the time the frequency drops to $\unit[80]{\%}$ of the initial frequency, the power will drop to $\unit[2.8]{\%}$ of the initial power. 
Therefore, a $\unit[0.8\times10^3]{Hz}$ signal, will have radiation power $ \sim 36$ times less than the power of a $\unit[1.0\times10^3]{Hz}$ signal and therefore, 
a signal to noise ratio (SNR) 36 times smaller than the SNR of a signal at $\unit[1.0\times10^3]{Hz}$. This means that we are more interested in the initial
stages of the spin-down. \\

The duration of the signal when frequency drops from $\unit[1.0\times10^3]{Hz}$ to $\unit[0.8\times10^3]{Hz}$ ranges from about $\unit[3.5\times10^2]{s}$ for 
$\alpha=1$ to about $3.5\times10^6$ for $\alpha=10^{-2}$ (derived using \eqref{eq.8}). Because of its duration, such a signal is classified as a long-transient 
gravitational-wave signal. For this reason gravitational-wave searches for the r-mode signals are suitable for cross-correlation type analyses 
\citep{STAMPPAPER}. \\

From \eqref{eq.16}, we see that for an initial frequency of $\unit[1.0\times10^3]{Hz}$ the r-mode gravitational-wave power ranges from order $\unit[\sim10^{46}]{ergs\,s^{-1}}$ for $\alpha= 10^{-2}$, 
to order $\unit[\sim10^{50}]{ergs\,s^{-1}}$ for $\alpha=1$. These values are comparable to the power of the f-mode gravitational radiation, which is the most energetic of all 
neutron star oscillation modes. Using the energy stored in the neutron star f-mode as given in \citep{FMODE}, with order ranging from $\unit[10^{48}]{ergs}$ to $\unit[10^{50}]{ergs}$
and also using the damping times of order $\unit[1.0\times10^2]{s}$, \citep{FERRARI} the gravitational-wave power of these modes is of order $ \unit[10^{46}]{ergs\,s^{-1}}$ to $\unit[10^{48}]{ergs\,s^{-1}}$, 
thus concluding that r-mode gravitational radiation may be (or better, it used to be thought of) as powerful as the neutron star's most powerful gravitational radiation. \\

Note that \eqref{eq.16} cannot be taken simply by taking the time derivative of \eqref{eq.12}. This is because expression \eqref{eq.12} 
is just the energy of the r-mode, and does note include the rotational energy of the neutron star itself. In the saturated phase 
of the r-mode evolution described by \eqref{eq.f}, it is assumed that non-linear fluid forces are coupling the r-mode to other modes of 
the star in such a way that gravitational radiation is able to extract energy from the overall rotational energy of the star, not 
just from the r-mode.  So during this phase it would not be appropriate to use \eqref{eq.12} as the total reservoir of energy that is being
radiated into gravitational waves. \\

\section{R-mode sources}

In this section we use the power dependence on $f$ and $\alpha$ as expressed in \eqref{eq.16}, to argue that the newborn neutron stars are the most promising sources
of detectable r-mode gravitational waves. We also discuss the relevant types of electromagnetic triggers as well as their event rates. \\

\subsection{Comparing newborn neutron stars to other sources}
Apart from newborn neutron stars there are several other sources of r-mode gravitational radiation: starquakes on isolated neutron stars \citep{DUNCAN}, pulsar postglitch 
relaxation \citep{VAHID,ROSEN}, accreeting low mass x-ray neutron stars (LMXB) in a binary system \citep{SDNS2007,LEVIN}, and oscillations of the remnant 
star (delayed collapse) during the merging phase (before collapsing to a black hole) of a compact binary coalescence (CBC) \citep{SHAPIRO,SHIBATA}. From \eqref{eq.16}
we see that, due to their high angular velocities newborn neutron stars will emit the most powerful r-mode gravitational radiation out of all of the above sources. 
Even if we consider all the CBC as r-mode radiation sources (ignoring the low likelihood of a delayed collapse) the event rate
is much lower than the supernova event rate (within the detection distances). The LMXB, the starquakes on neutron stars and the pulsar glitches may be more frequent 
events but the gravitational radiation emitted by these sources is much weaker than that (predicted to be) emitted by newborn neutron stars. \\

\subsection{Electromagnetic Counterparts}

The r-mode search from newborn neutron stars depends on electromagnetic triggers from supernova type-I and type-II explosions.  
From \eqref{eq.15} we see that a r-mode detection will give an estimate for the ratio $\alpha /d$. Therefore, to extract any information 
about the magnitude of $\alpha$ it is necessary to know the distance to the source. Distances to type-I supernova can be calculated using the standard 
candle method with an error between $\unit[5-10]{\%}$ \citep{TYPE1DIST}. Distances to type-II supernovae can be calculated using the expanding photosphere method 
giving an error of $\unit[10-15]{\%}$ \citep{photospheric,TYPEIIDIST}. Furthermore, our sensitivity study results (table 1 in section 11.1) show that aLIGO can be sensitive to r-mode 
signals only from newborn neutron stars within our local group of galaxies. Since distances to galaxies in our local group are already known a supernova
explosion within our local group would automatically give information about the distance to the hypothetical r-mode gravitational radiation source. \\

\subsection{Event rates within the detection distances}

Considering both supernova types together we expect $2-3$ galactic events per century \citep{galacticR}, 1 event every 2-3 years at a distance of $\unit[5]{Mpc}$ 
and 1-2 events per year at a distance of $\unit[10]{Mpc}$ \citep{allrates}. However, the local supernova rates seem to be higher around Milky Way: In the years
from 2002-2005 4 events were detected (1-2 expected) within $\unit[4]{Mpc}$ and 9 events were detected (4-8 expected) within $\unit[10]{Mpc}$. This information suggests 
that in the Milky Way's neighborhood the supernova rates are higher than the predicted ones, meaning that the supernova rates around Milky Way's local volume 
are higher than the rates at distant typical volumes. This is a motivation to improve our detection algorithms thus increasing the detection distances and 
cover as much of our local group of galaxies as possible. \\

The latest supernova (SN2014J) occurred in January of 2014 in the galaxy Messier 82 (M82) in the nearby group of galaxies M81 and it is a type-I supernova. 
This galaxy is at a distance of $\unit[3.5]{Mpc}$ from the Earth. Estimates show a supernova rate in our Local group of galaxies (up to a distance of $\unit[1.5]{Mpc}$ 
from Earth) of 3-6 per century \citep{SNRATE,MLG}. \\

\section{Time frame for a r-mode gravitational wave detection}

 In this section we discuss the physics of the neutron star matter via the use of the r-mode instability window to identify the time-frame for a hypothetical 
 r-mode detection. An electromagnetic trigger will not necessarily coincide with the emission of r-mode gravitational waves.
 Depending on the physics of the r-modes (saturation amplitude and dissipation forces) and also the cooling mechanism of the neutron star (that depends on the EOS 
 of the neutron star matter) the r-mode emission may start from several minutes to up to a year after an electromagnetic trigger. \\
 
\subsection{The r-mode instability survival window (temperature versus angular velocity)} 

 In a hot, newly born neutron star, the main dissipation mechanism is due to viscous forces \citep{RMODEINST,OSCINSTRELST}. For the simplest neutron star models, 
 two kinds of viscosity are normally considered: bulk viscosity and shear viscosity. At high temperatures (above $\unit[10^{10}]{^oK}$), bulk viscosity is the dominant dissipation 
 mechanism while shear viscosity dominates for temperatures below $\unit[10^6]{^oK}$. Hence, there is a window between $\unit[10^6-10^{10}]{^oK}$ where the r-mode instability may be active. \\

 The instability window is defined by the temperature interval in which the dominant dissipation mechanism of the r-mode oscillation is the emission of (r-mode) gravitational
 radiation.  \\  

 The amplitude, $A$, of the r-mode mass current oscillations evolves like 
 
 \begin{equation}
 \label{eq.17}
  A \sim \exp \left[ it \left( \Omega + \frac{i}{\tau} \right) \right] 
 \end{equation}

 \noindent
 where $\Omega$ is the angular velocity of the neutron star and $\tau$ is the (e-folding) time scale of the r-mode oscillations. The above expression can be expanded as
 
\begin{equation}
\label{eq.17b}
 A \sim \exp\left(i \Omega t\right) \exp\left(-\frac{t} {\tau_{g}}\right) \exp\left(-\frac{t}{\tau_{s}}\right) \exp\left(-\frac{t}{\tau_{b}}\right)
\end{equation}

\noindent
where $g$ stands for dissipation due to gravitational-wave emission, $s$ stands for dissipation due to shear viscosity and $b$ stands for dissipation
due to bulk viscosity. Equations \eqref{eq.17} and \eqref{eq.17b} imply that the imaginary part, $1/ \tau$, of the frequency is given by

\begin{equation}
\label{eq.19}
\frac{1}{\tau}= \frac{1}{\tau_{g}} + \frac{1}{\tau_{s}} + \frac{1}{\tau_{b}}
\end{equation}

\noindent
Assuming the r-mode mass perturbations, $\delta v$, have a time dependence that is given by \eqref{eq.17} 
then the perturbation energy has a time dependence like 

\begin{equation}
\delta v \delta v^* \sim \exp\left ( - \frac{2t}{\tau} \right)
\end{equation}

\noindent
Therefore, the time derivative of the perturbation energy is given by

\begin{equation}
\label{eq.20}
 \frac{dE}{dt}= - \frac{2E}{\tau}
\end{equation}

\begin{figure}
\epsscale{1.0}
\plotone{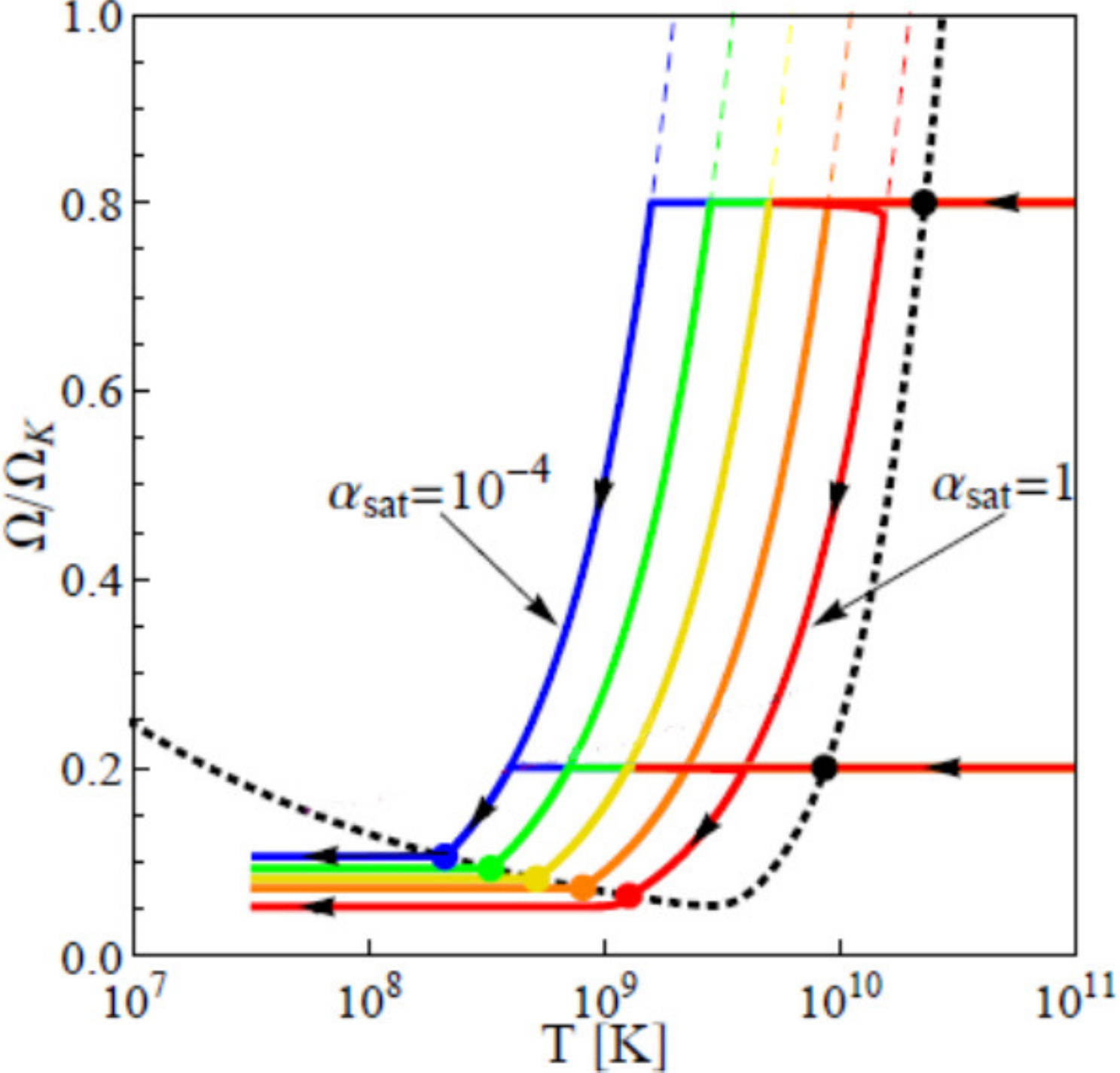}
\caption{Instability window of the r-mode: The neutron stars are born with temperatures of the order $\unit[10^{11}-10^{12}]{^oK}$. At high temperatures
  (above $\unit[10^{10}]{^oK}$) bulk viscosity is the dominant dissipation mechanism. Shear viscosity dominates for temperatures below $\unit[10^6]{^oK}$. 
  Hence there is a window between $\unit[10^6-10^{10}]{^oK}$ where the r-mode instability is active and can have an astrophysically significant role \citep{OSCINSTRELST}. 
  The dotted curve defines the boundary $\Omega_{critical}(T)$ of the instability region, the dashed curves (mostly hidden underneath the solid curves)
  represent the steady state, heating = cooling, and the solid lines show the numerical solution of the evolution equations for two fiducial initial
  rotational frequencies $\Omega=0.8 \Omega_K$ and $\Omega=0.2 \Omega_K$. As shown above, the higher the $\alpha$ value is the sooner the initial cooling phase
  will merge with the corresponding steady state curve. The above image is taken from \citep{ALFORD13}. } \label{Fig:ef2}
\epsscale{1.0}
\end{figure}

\noindent
and the r-mode instability condition can be expressed by  

\begin{equation}
\label{eq.21}
 \frac{1}{\tau} < 0 \,\,\,\,\,\,\,\,\ \mbox{or} \,\,\,\,\,\,\,\,\,\ \frac{1}{\tau_{g}} + \frac{1}{\tau_{s}} + \frac{1}{\tau_{b}} < 0
\end{equation}
 
 \noindent
 The conditions given by \eqref{eq.21} imply that the energy of the r-mode increases with time. Since $\tau_{s}$ and $\tau_{b}$ are always positive then $\tau_{g}$ has to be negative 
 and smaller than $\tau_{s}$ and $\tau_{b}$ such that 
 
 \begin{equation}
 \left| \frac{1} {\tau_{g}} \right| > \frac{1}{\tau_{s}} + \frac{1}{\tau_{b}}
 \end{equation}
 
\noindent 
for the r-mode instability to take place.
In general $1/ \tau$ is given in terms of the angular velocity of the neutron star $\Omega$ and its temperature T \citep{GWHYNS}. Thus, the equation 
$1/ \tau=0$ defines a curve of $\Omega_{critical}(T)$. Above this curve, the instability condition is satisfied, therefore, this $\Omega_{critical}(T)$ curve
defines the window over which r-mode instability will occur as seen in Fig.2. \\

 Bulk viscosity arises because of the pressure and density variations due to the r-mode oscillation driving the fluid away from beta equilibrium. Neutrinos carry 
 away the r-mode oscillation energy lost during this process. The balance between neutrino cooling and viscous heating plays an important role in the spin-down 
 evolution of a neutron star \citep{SDNS2009}. Also, the bulk viscosity can be strongly affected by the presence of hyperons in the neutron star core. In the 
 presence of hyperons the bulk viscosity coefficient is stronger and therefore, hyperon bulk viscosity becomes relevant at lower temperatures \citep{OSCINSTRELST}. \\ 
 
 Shear viscosity \citep{RMODEINST,OSCINSTRELST} is used to describe in a macroscopic manner the scattering events associated with momentum transfer. 
 Neutron-neutron scattering provides the most important contribution in a neutron star. However, in cases the core of the neutron star becomes superfluid,
 electron-electron scattering is the dominant effect that contributes to the shear viscosity. Therefore, a superfluid neutron star core can affect the 
 spin-down evolution. \\

 \subsection{Detection window after a supernova}

A neutron star is born with core temperatures of $\unit[\sim10^{12}]{^oK}$. Right after the catastrophic collapse, the formation of a protoneutron star is defined as the 
collapsed star with neutrinos trapped inside its core. About $\unit[1.0\times10^2]{s}$ after the collapse these neutrinos are released leading to a dramatic temperature drop to $\unit[10^{10}]{^oK}$
\citep{NSOEOS}. The transition from a neutrino-opaque protoneutron star core to a neutrino-transparent core, resulting in the neutrino burst observed
during a supernova, signals the death of the protoneutron star and the birth of a neutron star. \\

After the newborn neutron star enters the instability window, it undergoes an initial cooling. The spin-down starts after the steady state (cooling=heating) curve is reached. 
Depending on the saturation amplitude and the EOS, this can be reached at temperatures anywhere between $10^9-10^{10}$ \citep{ALFORD13}. That means the spin-down may start right after 
the neutrino burst ($\unit[1.0\times10^2]{s}$ after the collapse of the star) or up until the neutron star cools down to $\unit[10^9]{^oK}$. This time depends on the cooling rate of the neutron star. \\ 

The modified Urca process (MUP) can decrease the temperature of the neutron star from $T=\unit[10^{10}]{^oK}$ to $\unit[10^9]{^oK}$ in about a year and down to $\unit[10^8]{^oK}$ within a 
million years. However, if the direct Urca process (DUP) is doing the cooling, the core temperature will drop to $\unit[10^9]{^oK}$ within minutes and then drop down 
to $\unit[10^8]{^oK}$ within days \citep{NSC}. The EOS determines which cooling evolution, $T(t)$, the neutron star will follow. For a given EOS, the transition rate from slow to fast 
cooling occurs in a given mass range. This range is very narrow because of the sharp neutron core mass density threshold for the direct Urca process to take place \citep{NSCS}. \\ 
  
The growth timescale of the r-mode after it enters the instability window is of the order of $\unit[\sim 40]{s}$. Therefore, within $10$ minutes the amplitude may grow up to order 
of $1$ if not saturated before that. As saturation is reached the r-mode gravitational radiation will equilibrate at a particular heating=cooling curve (Fig.\ref{Fig:ef2}). 
How long after the birth of the neutron star will this take place? From the above paragraphs we see that, depending on the equation of state and the r-mode 
saturation amplitude, the spin-down of a newborn neutron star may start anywhere from a few minutes up to $ \sim$years after the neutron star is born. Therefore we conclude that 
if aLIGO takes data in 2015, it may be possible to set some constraints for the EOS of the neutron star remnant of the Messier 82 supernova (SN2014J) that occurred in January 2014. \\

\section{Motivation for the r-mode sensitivity study: MOI of a neutron star from a r-mode gravitational wave detection}
 
In this section we derive a generalization of the r-mode waveform \eqref{eq.6} so that the frequency evolution is expressed in terms of any EOS. 
We then derive an expression for the MOI of the neutron star as a function of the r-mode gravitational-wave observables, $\dot E$, $\dot f$ and $f$. 
Subsequently we show how the value of the MOI sets constraints on the possible EOS of the neutron star matter. This result is used in section 7 where 
we derive an expression for the saturation amplitude, $\alpha$, as a function of the EOS. Using this expression, a set of possible values of $\alpha$ 
can be derived, thus setting an upper bound for $\alpha$. \\ 

\subsection{Estimating the moment of inertia of a neutron star from a hypothetical r-mode detection}

In section 2 we have assumed a polytropic EOS for the neutron star matter. The constant $\mu$ in \eqref{eq.7} as well as the waveforms expressed in \eqref{eq.6}
are dependent on this EOS. Therefore, a hypothetical r-mode detection would set upper bounds on the r-mode saturation amplitude $\alpha$ based on the 
assumed EOS. Using the same hypothetical detection two different EOS may result in two different upper bounds for $\alpha$. \\

In what follows, we derive an EOS-dependent and $\alpha$-independent expression \eqref{eq.45} or \eqref{eq.46} (which turns out to be the moment of inertia 
of the neutron star) that can be evaluated from a r-mode detection. Thus we conclude that a hypothetical r-mode detection can constrain the EOS of the neutron
star matter and can also set upper bounds on the values of $\alpha$ for each permitted EOS. \\ 

In this section we seek a generalization of the spin-down formula in \citep{GWHYNS}, so that no assumptions are made for the equation of state (EOS) of the neutron star matter.
The gravitational radiation time scale is given by

\begin{equation}
\label{eq.31}
 \frac{1}{\tau_{gw}} = - \lambda_m M R^{2m}\Omega^{2m+2} \tilde{J_m}
\end{equation}

\noindent
where 

\begin{equation}
\label{eq.32}
 \lambda_m =  \frac{32 \pi G }{c^{2m+3}} \left ( \frac{(m-1)^{m}}{(2m+1)!!} \right)^2 \left( \frac{m+2}{m+1} \right)^{2m+2}
\end{equation}

\noindent
and

\begin{equation}
\label{eq.33}
 \tilde{J_m}= \frac{1}{MR^{2m}} \int^R_0 \rho(r) r^{2m+2} dr
\end{equation}

\noindent
The energy $ E_m$ of the r-mode is given by
\begin{equation}
\label{eq.34}
 E_m = \frac{1}{2} \alpha^2 \Omega^2 MR^2 \tilde{J}_{m}
\end{equation}

\noindent
The rate of change of the energy of the r-mode oscillation is given by

\begin{equation}
\label{eq.35}
 \dot{E}_m= - \frac{2E}{\tau_{gw}}
\end{equation}

\noindent
and hence from \eqref{eq.31}, \eqref{eq.34} and \eqref{eq.35} we obtain

\begin{equation}
\label{eq.36}
 \dot{E}_m = \lambda_m \alpha^2 M^2 R^{2m+2}\Omega^{2m+4} \tilde{J}^2_m
\end{equation}

During the saturated non-linear phase of the evolution, when the r-mode oscillation has a fixed amplitude and energy, \eqref{eq.36} gives the 
power of the mode being radiated as r-mode gravitational radiation. During this phase the main contribution to the r-mode is the $m=2$ harmonic 
and the angular velocity of the star evolves according to \citep{GWHYNS} \\

\begin{equation}
\label{eq.37}
 \dot{\Omega} = \frac{2 \Omega}{\tau_{gw}} \left ( \frac{\alpha^2 Q}{1- \alpha^2 Q} \right ) \approx \frac{2 \Omega \alpha^2 Q}{\tau_{gw}}
\end{equation}

\noindent
where the last step assumes that $\alpha^2 \ll 1$ and Q is defined by

\begin{equation}
\label{eq.38}
 Q= \frac{3 \tilde{J}_2}{2 \tilde{I}}
\end{equation}

\noindent
where

\begin{equation}
\label{eq.39}
 \tilde{J}_2= \frac{1}{MR^{4}} \int^R_0 \rho(r) r^{6} dr 
\end{equation}

\noindent
and 

\begin{equation}
\label{eq.40}
\tilde{I}= \frac{8 \pi}{3MR^{2}} \int^R_0 \rho(r) r^{4} dr 
\end{equation}

\noindent
Substituting \eqref{eq.31}, \eqref{eq.32} and \eqref{eq.38} in \eqref{eq.37} (for $m=2$) results in the EOS-dependent expression for the time evolution of the angular velocity of the neutron star

\begin{equation}
\label{eq.41}
 \dot{\Omega} = - 3 \lambda_2 \alpha^2 M R^4  \frac{\tilde{J}_2^2} {\tilde{I}} \Omega^7
\end{equation}

\noindent
where $\lambda_2= 7.7\times 10^{-70}$. Using the result relating the angular velocity, $\omega=2\pi f$, of the r-mode gravitational radiation to the angular 
velocity, $\Omega$, of the neutron star $\omega = 4/3 \times \Omega \implies  f= 2 /3\pi \times \Omega$, we get the gravitational radiation frequency evolution given by

\begin{equation}
\label{eq.42}
 \dot{f} = - 2.5 \times 10^{-65} M R^4 \alpha^2  \frac{(\tilde{J}_2)^2}{\tilde{I}} f^7 \text{kg}^{-1} \text{m}^{-4} \text{s}^5
\end{equation}

\noindent
leading to the EOS-dependent r-mode radiation waveform 

\begin{equation}
\label{eq.43}
  f = \left( \frac{1}{ f_o^{-6} + \mu t } \right) ^{ \frac{1}{6} }
\end{equation}

\noindent
where

\begin{equation}
\label{eq.44}
 \mu = 1.5 \times 10^{-64} M R^4 \alpha^2 \frac{( \tilde{J}_2 )^2}{\tilde{I}} \text{kg}^{-1} \text{m}^{-4} \text{s}^5
\end{equation}

\noindent
and $f_o$ is the initial frequency of the neutron star spin-down. \\

Using \eqref{eq.36} for $m=2$, and \eqref{eq.41} we get the result

\begin{equation}
\label{eq.45}
 \frac{\dot{E}}{\Omega \dot{\Omega}} = -\frac{8 \pi}{9} \int^R_0 \rho(r)r^4 dr
\end{equation}

\noindent
or, in terms of the gravitational-wave frequency, $f$, 

\begin{equation}
\label{eq.46}
 \frac{\dot{E}}{f \dot{f}} = -2 \pi^3 \int^R_0 \rho(r)r^4 dr
\end{equation}

\noindent
A hypothetical r-mode detection will provide us with estimates of the model parameters $f_o$ and $\mu$. From these we can estimate the r-mode gravitational 
wave frequency, $f$, as well as its derivative, $\dot{f}$. Furthermore, from equation \eqref{eq.13} we see that given the distance, $d$, to the source of 
the r-mode gravitational radiation, the frequency, $f$, of the gravitational wave and the gravitational wave strain, $h$, then the power, $\dot{E}$, 
of the r-mode gravitational radiation can be estimated. Therefore, we conclude that given a hypothetical r-mode gravitational wave is detected, the 
integral on the right hand side of \eqref{eq.46} that gives the moment of inertia of the neutron star can be estimated. \\

\section{Constraining the EOS of neutron star matter and the r-mode saturation amplitude}

In this section we plot the moment of inertia versus radius and the moment of inertia versus mass representations of 17 EOS. From a 
hypothetical r-mode detection we can estimate the moment of inertia of the neutron star and hence estimate the possible radius and mass 
values for the allowed EOS. Subsequent estimates of either the radius or mass of the neutron star would further constrain the EOS. For each possible 
EOS we can then set upper bounds on the r-mode saturation amplitude, $\alpha$. We also discuss the clues an r-mode gravitational wave detection will 
provide about the cooling mechanism of the neutron star. \\

\subsection{Numerical solution to the TOV equations}

Given a pair $(\rho(P),P_c)$ of an EOS and a central pressure of a neutron star we can integrate the Tolman-Oppenheimer-Volkov (TOV) equations

\begin{equation}
\label{eq.50}
 \frac{dP}{dr} = - \frac{ G \left ( \rho(P)+\frac{P}{c^2} \right )  \left ( m+\frac{4 \pi r^3 P}{c^2} \right ) }{r \left ( r-\frac{2Gm}{c^2} \right)}
\end{equation}

\noindent
and 

\begin{equation}
\label{eq.50b}
 \frac{dm}{dr}= 4 \pi r^2 \rho
\end{equation}

\noindent
from $r=0..r'$ and $P=P_c..P(r')$. At every integral step we evaluate the following integrals

\begin{equation}
\label{eq.In}
 I_n(r') = \int^{r'}_0 \rho(P) r^n dr \,\,\,\,\,\,\, \mbox{with n=2,4 and 6} 
\end{equation}

\noindent
which can be used to re-write \eqref{eq.40} and \eqref{eq.39} like

\begin{equation}
\label{IJ}
\tilde{I}=\frac{8 \pi}{3MR^2}I_4(R) \,\,\,\,\,\, \mbox{and} \,\,\,\,\,\, \tilde{J}_2=\frac{1}{MR^4}I_6(R)
\end{equation}

\noindent
respectively, as well as \eqref{eq.46} like

\begin{equation}
 \frac{\dot{E}}{f \dot{f}} = - 2 \pi^3 I_4(R)
\end{equation}

\noindent
where $R$ is the value of $r'$ for which

\begin{equation}
 \label{eq.R}
 P(r')=0
\end{equation}

\noindent
Using equation \eqref{eq.In} with $n=2$ and $r'=R$ we can express the total mass of a neutron star of radius $R$ by

\begin{equation}
 \label{mass}
 M(R)=4 \pi I_2(R)
\end{equation}

\begin{figure}
\epsscale{1.3}
\plotone{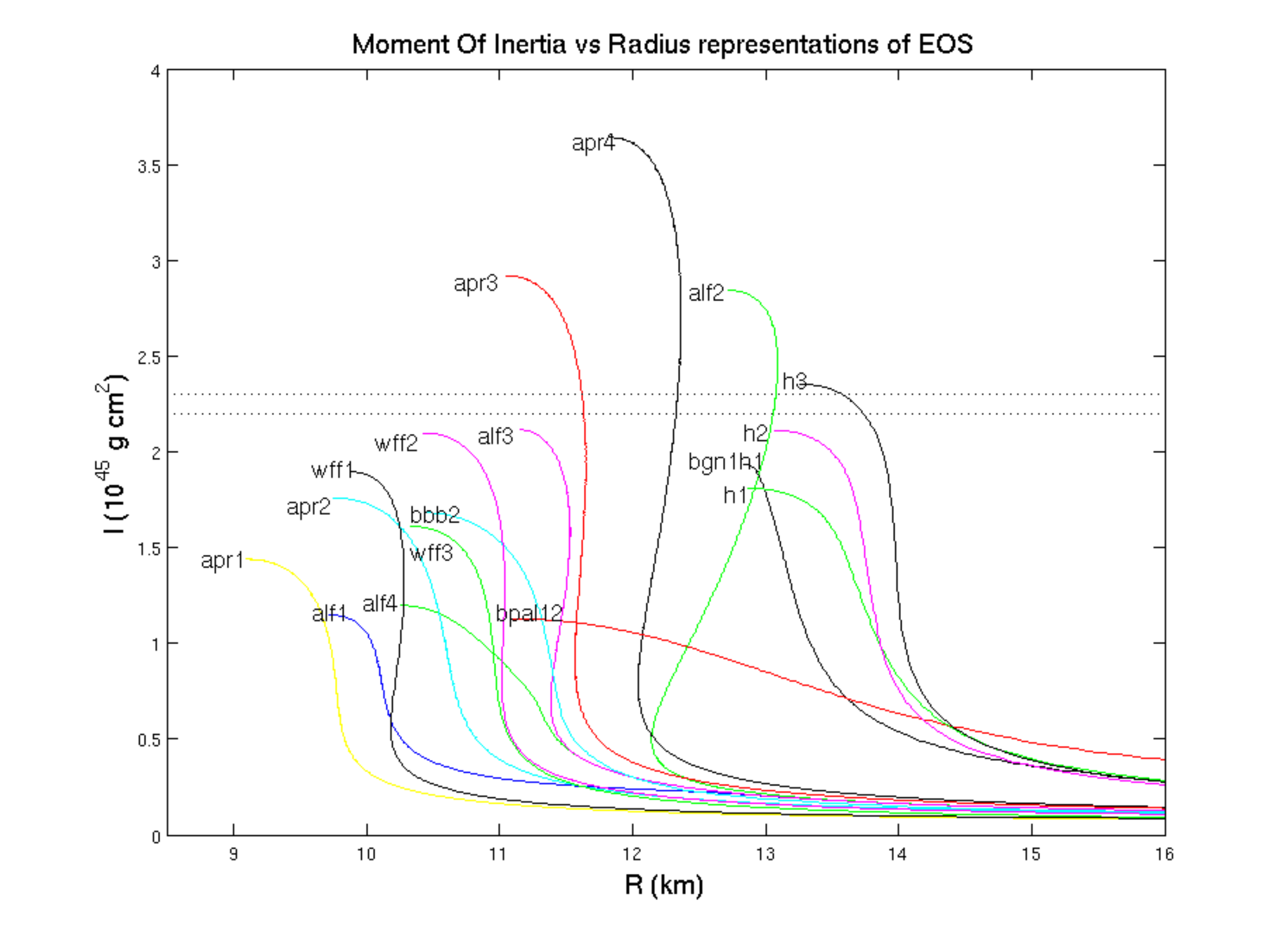}
\caption{Moment of inertia versus radius representation of the EOS. The horizontal lines correspond to a range of values of the moment of inertia
(estimated from a hypothetical r-mode detection), in this case $I_1 = 2.2$ and $I_2=2.3$. The corresponding radius range is shown in Fig.\ref{Fig:ef10}.  
} \label{Fig:ef8}
\epsscale{1.3}
\end{figure}

\begin{figure}
\epsscale{1.3}
\plotone{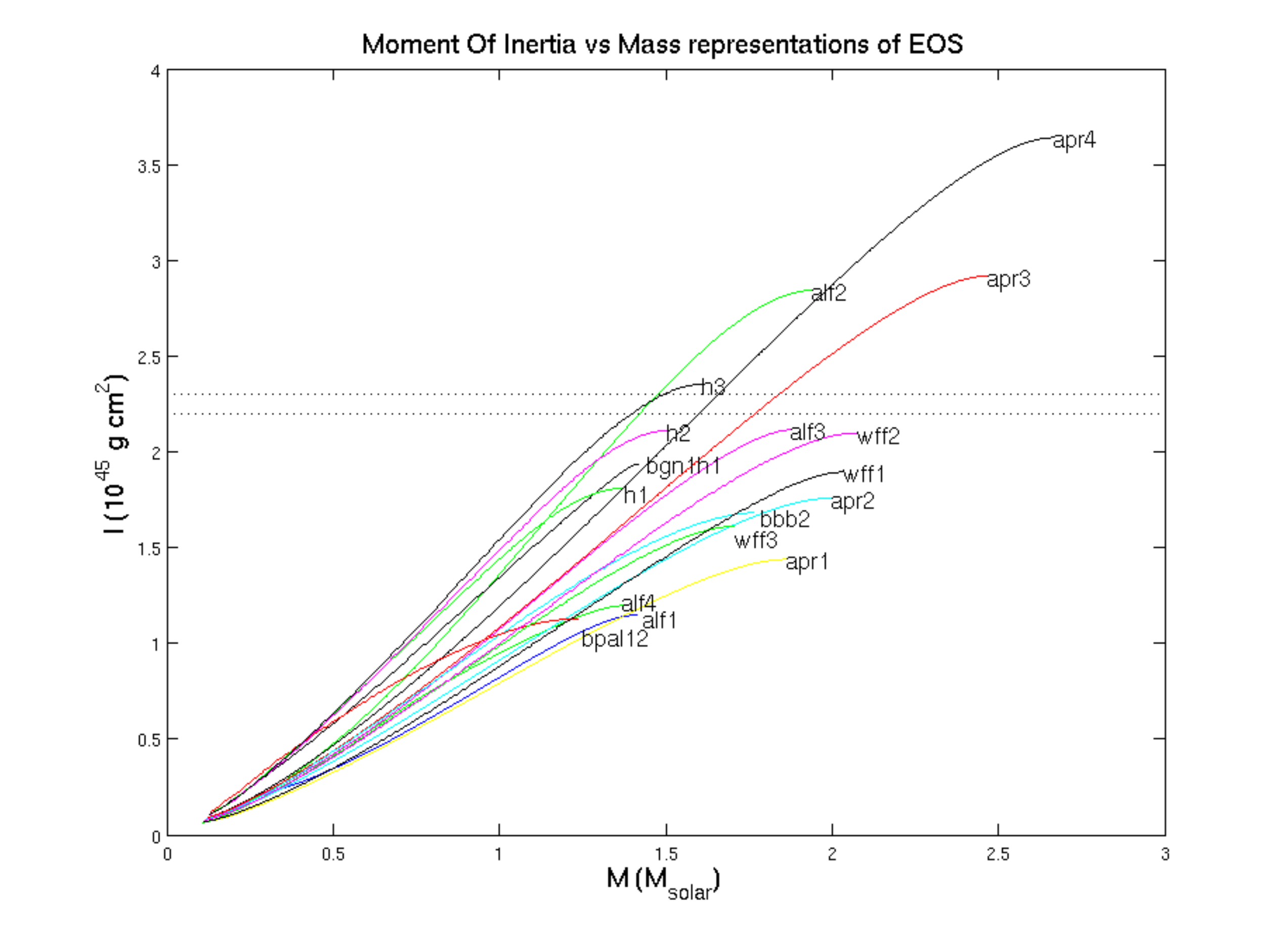}
\caption{Moment of inertia versus mass representation of the EOS. The horizontal lines correspond to a range of values of the moment of inertia
(estimated from a hypothetical r-mode detection), in this case $I_1 = 2.2$ and $I_2=2.3$. The corresponding mass range is shown in Fig.\ref{Fig:ef10}. } \label{Fig:ef9}
\epsscale{1.3}
\end{figure}

\begin{figure}
\epsscale{1.3}
\plotone{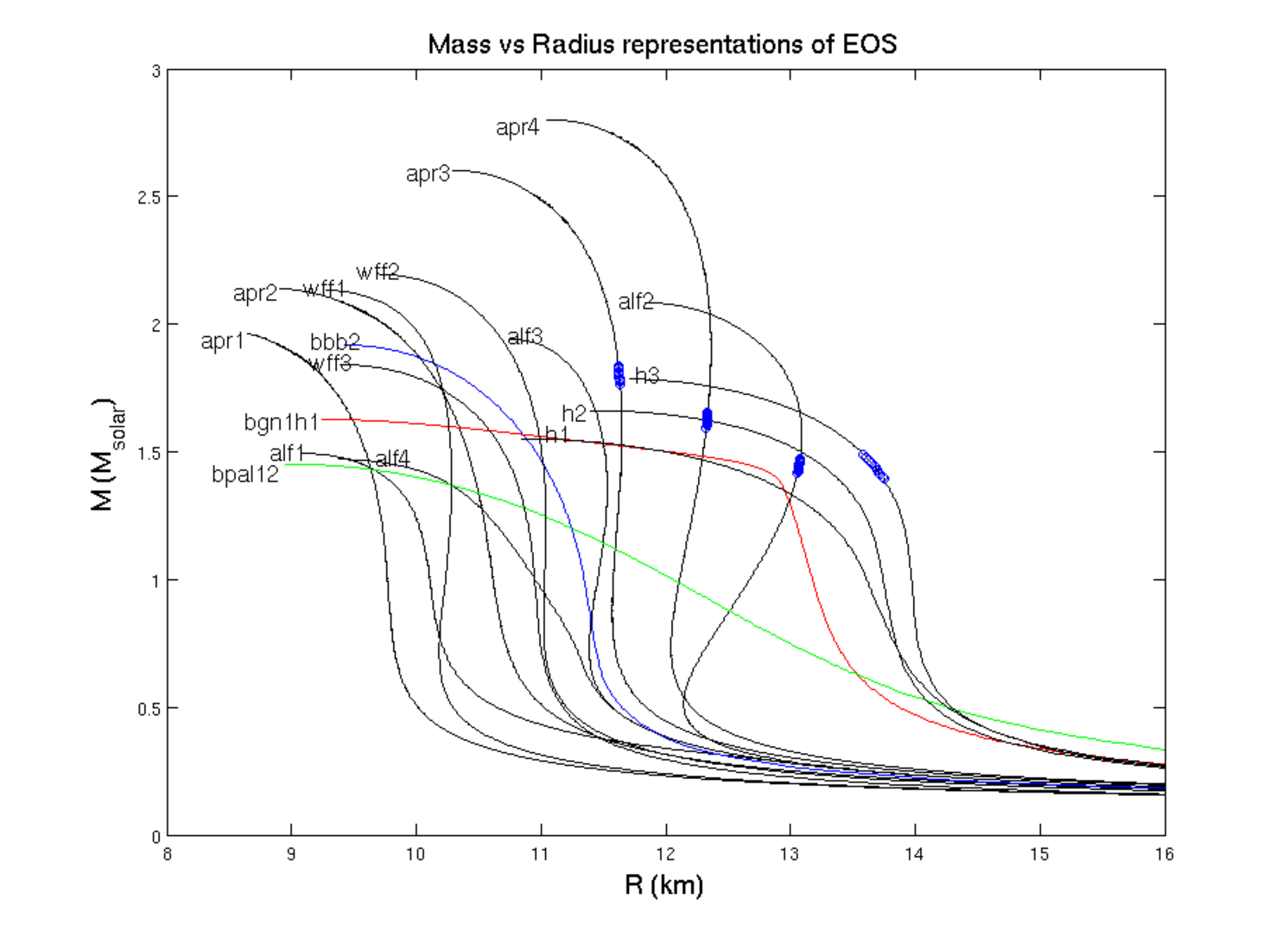}
\caption{Mass versus radius representation of the EOS. Each value of the moment of inertia determined by a hypothetical r-mode detection corresponds to a point
on one (or more) $M$-$R$ curves. A range of moment of inertia values would correspond to a part of the $M$-$R$ curves. Depending on what EOS curves are cut by the horizontal lines in 
Fig.\ref{Fig:ef8} and Fig.\ref{Fig:ef9} constraints may be put on what the possible EOS are for the neutron star matter. } \label{Fig:ef10}
\epsscale{1.3}
\end{figure}

Solving \eqref{eq.50}, \eqref{eq.50b} and \eqref{eq.In} (for the given $\rho(P)$) while varying $P_c$, results in the functions 
$R=R(P_c)$, $I_2(R,P_c)$ and $I_4(R,P_c)$ where $I_2$ and $I_4$ are proportional to the neutron star mass and moment of inertia respectively. 
Using these results we can express the total mass, $M$, and the moment of inertia, $I$, of the neutron star as $M(P_c)$ and $I(P_c)$.
Expressing the macroscopic observables, I, R and M in terms of the central pressure, $P_c$, demonstrates that the equation of state of the neutron 
star matter can be thought of as: (a) a curve in the $I-R$ plane (Fig.\ref{Fig:ef8}), (b) a curve in the $I-M$ plane (Fig.\ref{Fig:ef9}) and 
(c) a curve in the $M-R$ plane (Fig.\ref{Fig:ef10}); all curves being parametrized by $P_c$, with each $P_c$ value corresponding to a unique 
point on each curve. Figures 3, 4 and 5 were plotted by integrating the TOV equations using the Runge-Kutta method for a set of 17 equations
of state.  \\

A hypothetical r-mode gravitational-wave detection, with r-mode waveform parameters $f_o$ and $\mu$, will give an estimate of the moment of inertia, 
$I$ ($\propto I_4$), of the neutron star and hence intersect the corresponding $I-R$ and $I-M$ curves. The points of intersection represent the set
of all possible mass and radius values allowed by the estimated value of $I$. The mass and radius values are shown in (Fig.\ref{Fig:ef8}) and (Fig.\ref{Fig:ef9}). 
Using these figures we see that a subsequent measurement of either the mass or the radius of the neutron star, would pick a single $I-R$ or $I-M$ 
curve and hence fix a choice of the EOS. The pairs of all possible $(M_i,R_i)$ points ($i$ corresponding to each 
intersection point in (Fig.\ref{Fig:ef8}) and (Fig.\ref{Fig:ef9})) are shown in (Fig.\ref{Fig:ef10}) which is the standard $M-R$ representation of the EOS. \\

Upon constraining the EOS to a few options we can substitute the corresponding values of $\tilde{I}$, $\tilde{J}_2$ and $R$ as given by \eqref{IJ} and \eqref{eq.R} 
respectively in \eqref{eq.44} to get 

\begin{equation}
\label{eq.52}
 \mu = 3.0 \times 10^{-66} \frac{\alpha^2}{R^2} \frac{\left( I_6(R) \right)^2}{\left( I_4(R) \right)}
\end{equation}

\noindent
Rearranging \eqref{eq.52} we can express $\alpha$ as a function of the value of $\mu$ estimated from the hypothetical r-mode detection and the allowed EOS as follows

\begin{equation}
 \label{eq.53}
 \alpha = 5.8 \times 10^{32}  \frac{R \sqrt{\mu I_4(R)} }{I_6(R)}
\end{equation}

\noindent
Therefore, we can set constraints (for each allowed EOS) on the r-mode saturation amplitude. \\

\subsection{EOS, $\alpha$ and the cooling rate of the neutron star}

After the protoneutron star core becomes transparent to neutrinos, the cooling mechanism (and hence the time at which the neutron star hits the heating=cooling curve) depends 
on the EOS of neutron star matter. For non-superfluid neutron stars there are two distinct and well known cooling mechanisms: fast neutrino cooling via the direct Urca process 
(DUP) and slow neutrino cooling via the modified Urca process (MUP). For a given EOS there is a density threshold, $\rho_{th}$, beyond which DUP can occur. For the given EOS, 
there is a total mass threshold, $M_{th}$, required for the central density $\rho_c$ to be equal to $\rho_{th}$. Neutron stars of total mass below $M_{th}$ can only cool through MUP. \\

The cooling evolution splits into two universal categories: the (slow) cooling for neutron stars of mass less than the threshold mass, $M_{th}$, and the (fast) cooling 
for neutron stars of mass greater than the threshold mass, $M_{th}$. The cooling curves in each category are universal i.e. they are independent of the mass or EOS. \\

Assume the neutron star is born with an angular velocity $\Omega_o$ that stays approximately constant (ignoring magnetic braking) until the r-mode gravitational radiation 
starts spinning down the neutron star. A hypothetical r-mode detection will provide us with the value of $\Omega_o$ as well as with the value of $\mu$ as shown in \eqref{eq.52}. 
This value of $\mu$ together with the chosen EOS can be used to  determine the saturation amplitude $\alpha$ which in turn can be used to determine the heating=cooling curve 
as shown in Fig.\ref{Fig:ef2}. The horizontal line $\Omega=\Omega_{o}$ intersects the heating=cooling curve at $T=T_r$ and hence $T_r$, the temperature at which the neutron 
star starts spinning down due to r-mode gravitational radiation, can also be estimated. \\

The time, $t_r$, at which the neutron star temperature is equal to $T_r$ can be estimated by the initial detection time of the r-mode gravitational radiation. 
Furthermore, the time, $t_o$, at which the neutron star is born is given by the time at which the neutrino burst occurs and may also be detected. Since 
the temperature of the neutron star at birth is generically approximately equal to $\unit[10^{10}]{^oK}$, $t_o$ gives the time at which the neutron star 
has this temperature. \\

The time elapsed between $t_r$ and $t_o$ provides the time taken for the neutron star temperature to drop from $\sim \unit[10^{10}]{^oK}$ to $T_r$. Therefore, the 
neutron star cooling rate can be estimated. This cooling rate may be consistent either with DUP (total mass has to be above $M_{th}$) or MUP (total mass has to be 
below $M_{th}$). If the cooling mechanism (implied by the cooling rate) is not consistent with the mass of the neutron star (implied by the chosen EOS) that means 
the chosen EOS is not the correct one. \\

Constraining the EOS of a neutron star using cooling mechanism considerations as well as the value of the MOI of the neutron star (as predicted by a hypothetical r-mode 
detection) together with equations \eqref{eq.52} and \eqref{eq.53} are the main results of the first part of the paper. These results demonstrate the importance of 
a hypothetical r-mode detection as a source of information about the physics of the state of matter in the interior 
of the neutron star as well as the neutron star mass-current oscillations. \\

This information carried by r-mode gravitational waves constitutes the motivation for the second part of the paper: to perform a sensitivity study on r-mode signals from newborn 
neutron stars. This study starts with section 8, where we first determine the characteristics of a set of representative waveforms that were injected on background detector noise and then 
recovered using a clustering algorithm. \\

\section{Waveform parameters: range of possible values of $\MakeLowercase{f}_o$}

In the sections that follow we are performing a sensitivity study to examine the distances
at which the aLIGO and ET detectors (combined with the preprocessing and decision making algorithms) will be sensitive for such signals. 
To perform the sensitivity study we need to choose a representative set of waveforms whose choice depends on the range of values of the waveform parameters, $f_o$ and $\alpha$.
To cover the whole spectrum of possibilities we need to determine the highest theoretically possible rotational velocities of neutron stars at birth and also determine the 
highest theoretically possible values of $\alpha$. In this section we determine the possible range of values for the rotational velocity of neutron stars at birth.  \\ 

\subsection{Theoretical upper bound for the neutron star angular velocity at birth}

The observed neutron star angular velocities are up to 25 times less than the theoretical upper bound of angular velocities that neutron stars may possess at birth.
Conservation of angular momentum of the progenitor star results in much higher angular velocities at birth than those our observations reveal.
Therefore, there must be a mechanism that causes the neutron star to spin down to the observed angular velocities. \\

 A progenitor mass between 8 and 25 solar masses will result in a supernova with a neutron star as a remnant. Numerical evolutions \citep{HegEtal}
 have shown that such progenitors reach angular velocities very close to the critical ones (called Kepler angular velocities) at which the stars would fall apart 
 due to centrifugal forces. Three different approaches (general relativistic with and without differential rotation as well as numerical evolution of a pre-supernova 
 stellar model) point towards an upper bound of $\unit[7.2\times10^3]{rad\,s^{-1}}$ for the critical angular velocity of such neutron stars. In the Newtonian approach, this limit 
 is obtained using
 
 \begin{equation}
 \label{eq.1}
  v_{escape}=\sqrt{\frac{2GM}{R}} 
 \end{equation}

 \noindent
 where $G$ is the gravitational constant, $M$ is the mass of the neutron star and $R$ is the radius of the neutron star.
 Equation \eqref{eq.1} gives 
 
\begin{equation}
\label{eq.2}
 \Omega_{Kepler_{(N)}}= 1.6 \sqrt{\pi G \tilde{\rho}}
\end{equation}

 \noindent
 while general relativistic corrections give 
 
 \begin{equation}
\label{eq.2}
 \Omega_{Kepler_{(GR)}}= \frac{2}{3} \sqrt{\pi G \tilde{\rho}}
\end{equation}
 
  \noindent
 Assuming a solid neutron star without differential rotation of uniform average density $\tilde{\rho}_{ns}=\unit[5.5 \times 10^{17}]{ kg\,m^{-3}} $ 
 we find that newborn neutron stars can reach angular velocities of $ \Omega_{ns}= \unit[7.2\times 10^3]{rad\,s^{-1}}$ 
 without falling apart due to centrifugal forces. \\
 
 In general relativistic models with differential rotation \citep{DimmOtt} the iron cores of supernova progenitor stars may rotate at angular velocities 
 in the range of $\unit[0.45-13]{rad\,s^{-1}}$ ($\Omega_{Kepler_{(GR)}} \simeq \unit[6.7]{rad\,s^{-1}}$ for an iron core with mass $M=1.4$ solar 
 masses and radius $R=\unit[1.0\times10^3]{km}$). Using conservation of angular momentum we get

 \begin{equation}
 \label{eq.3}
 \Omega_{ns} =  \left(   \frac{M_{ic}}{M_{ns}}  \right)  \left( \frac{R_{ic}}{R_{ns}}    \right)^2  \Omega_{ic}
 \end{equation}
 
 \noindent
 From \citep{DimmOtt} we find that the iron cores of supernova progenitors have masses in the range of $1.2-1.8$ 
 solar masses and radii from $\unit[1.0\times10^3]{km}$ to $\unit[2.5\times10^3]{km}$. Substituting, $\Omega_{ic}= \unit[1.0]{rad\,s^{-1}}$, 
 $M_{ic} = M_{ns} = 1.4$ solar masses, $R_{ic}=\unit[1.0\times10^3]{km}$ and $R_{ns}=\unit[12]{km}$, we find that the angular velocity of a new-born neutron star 
 is about $\unit[6.9\times10^3]{rad\,s^{-1}}$. This is consistent with the $\unit[7.0\times10^3]{rad\,s^{-1}}$ upper limit found by Hashimoto in 1994 \citep{UPLANGV}.\\
 
 Simulating the angular momentum evolution of massive stars ($8-25$ solar masses) in their spherically symmetric models \text{Heger et al.\,\,}presented the first 
 rotating presupernova stellar models that included a 1-dim prescription for angular momentum transport and centrifugal effects \citep{Hegetal2}. These authors 
 estimated the neutron star angular velocity by assuming that the total angular momentum contained in the pre-collapse iron core is conserved during collapse 
 and supernova phases and is deposited completely in a rigidly rotating neutron star of radius equal to $\unit[\sim12]{km}$ and moment of inertia given by 
 $ 0.35 M_{ns} R^2$ \citep{SPINOTT,LatPrak}. In this way, they estimated an initial neutron star angular velocity of $ \unit[6.3\times10^3]{rad\,s^{-1}}$ for 
 their fiducial 20 solar masses (zero age main sequence) stellar model. \\

 \subsection{Possible mechanisms responsible for the neutron star spin-down}
 
 According to the observational data, the angular velocities of neutron stars range between $\unit[\sim1.0]{rad\,s^{-1}}$ to $\unit[\sim1.0\times10^4]{rad\,s^{-1}}$. 
 Typical (generic pulsar) neutron stars have angular velocities from $\unit[\sim20]{rad\,s^{-1}}$ to $\unit[\sim3.0\times10^2]{rad\,s^{-1}}$. The highest of these is 
 up to a factor of $24$ less than the theoretical upper bound of the initial angular velocity of the newborn neutron star. Several suggested processes 
 that could lead to an early spin-down of a newborn (proto)neutron star include \citep{SPINOTT}: \\

   \noindent
 (a) angular momentum redistribution by global hydrodynamic angular instabilities \\
 (b) r-modes and gravitational radiation back-reaction \\
 (c) rotation-powered explosions  \\
 (d) viscous angular momentum transport due to convection \\
 (e) neutrino viscosity or dissipation of shear energy stored in differential rotation  \\
 (f) magneto-centrifugal winds  \\
 (g) early magnetic dipole radiation  \\
 (h) late-time fall-back  and  \\
 (i) anisotropic neutrino emission \\

 It is also possible that slowly spinning cores are the natural end-product of stellar evolution. Angular momentum transport via magnetic 
 processes yields angular velocities equal to $\unit[\sim6.0\times10^{-2}]{rad\,s^{-1}}$ \citep{SprPhin} that is too low to explain the observational 
 data. The authors had to rely on subsequent spin-up by off-center birth kicks to obtain angular velocities of $\unit[\sim3.0]{rad\,s^{-1}}$ to match the 
 observations. \text{Heger et al.\,\,\,}followed a similar approach for magnetic angular momentum transport during stellar evolution to obtain neutron star angular 
 velocities of $\unit[\sim9.0\times10^2]{rad\,s^{-1}}$ \citep{HegEtal}. Having in mind that stellar evolution theories with rotation are still improving and given the 
 uncertainties in iron core angular velocities and angular momentum profiles we should be very careful in accepting any spin-down mechanism as the 
 final theory. \\

 In this paper we assumed that the neutron star is born with initial angular velocities of up to $\Omega_{ns} \unit[\simeq7.2\times10^3]{rad\,s^{-1}}$ and the neutron star 
 has a temperature range that allows the r-mode gravitational instability to survive. Since the rotational energy of the neutron star is the energy source of the r-mode 
 oscillations, the emission of r-mode gravitational radiation is assumed to be responsible for the decreasing rotational energy of the neutron star. \\  

\section{Waveform parameters: range of possible values of $\alpha$} 
 
 In this section we determine (by doing a thorough examination of the literature review) the highest theoretically permitted value of $\alpha$. 
 The value of $\alpha$ has been the subject of many debates since the earliest work on r-mode gravitational waves. Here we present several contradictory results 
 that were published during the last 15 years about the maximum possible value of $\alpha$. The still ongoing debate on the subject made our choice of range of 
 values of $\alpha$ a little tricky. \\
 
\subsection{Historic perspective of the r-mode saturation mechanism}  

 The hypothesis that newborn neutron stars may spin down due to r-mode gravitational radiation was based on three major discoveries: \\
 
 \noindent
 (a) The existence of a secular instability in rotating Jacobian ellipsoids describing an increase in the angular velocity due to gravitational radiation \citep{SOLTWOPRBL}. \\ 
 (b) The proof that non-axisymmetric modes of the form $e^{im\phi}$ ($\phi$ being the azimuthal angle) for $all$ rotating stars are driven towards instability by gravitational 
     radiation reaction \citep{LAGRANGEPERT,SECINSTROTNS}. This instability is known as the Chandrasekhar-Friedman-Schutz (CFS) instability. \\
 (c) The discovery of non-radial low frequency oscillation modes on white dwarfs similar to the Rossby waves (hence the name r-modes) in the Earth's oceans and atmosphere \citep{FIRSTRMODE}. \\ 

 Though some work on r-mode oscillations on neutron stars was done during the 1980's \citep{SAIO,WAGONER} it wasn't untill 1998 that it was demonstrated that neutron 
 star r-mode oscillations satisfy the requirements for the CFS instability \citep{GRINSTHYNS}. \text{Lindblom et al.\,\,}showed that r-modes belong to 
 the group of non-axisymmetric modes that can be driven unstable due to gravitational radiation. This theory predicted that (assuming the r-mode oscillation 
 amplitude grows sufficiently large) r-mode gravitational radiation (primarily in the $m=2$ harmonic) would carry away most of the angular momentum of a 
 rapidly rotating newborn neutron star in the form of r-mode gravitational radiation. \\

 In the early years after the discovery of the r-mode instability, initial scenarios assumed that the r-mode amplitude would grow to order unity \citep{NUMNONLINEVOL} before 
 some unknown process would saturate the mode \citep{GWRMRRS,RMODEINST}. However, there were speculations that some non-linear hydrodynamics of the star 
 might limit the growth of the r-mode to very small values. This could be the result of r-mode oscillations leaking energy (due to non-linear couplings) 
 into other inertial modes faster than gravitational radiation reaction force could restore it. Since 1998 several studies on the saturation mechanism 
 showed a variety of results. \\
 
 Early work \citep{SCSRM} showed that the hypothesis of r-mode gravitational radiation cannot be applied to cold ($\text{T} < \unit[10^8]{^oK}$) neutron stars whose crust is not in a fluid state. 
 It was shown that in neutron stars colder than $\unit[10^8]{^oK}$ the crust might be perfectly solid and that would completely suppress the r-mode instability. However, the question about the r-mode amplitude 
 growth and saturation on hot ($\text{T} > \unit[10^8]{^oK}$) neutron stars remained open. \\
 
 In the absence of any saturation mechanism the r-mode amplitude would grow to values of order $1$. This idea was supported by early relativistic simulations that showed that 
 on a fixed neutron star geometry no r-mode amplitude saturation occurs even at large amplitudes \citep{NONRELRMODES}. These findings were further reinforced by the results of
 fully non-linear, 3-dim numerical simulations based on Newtonian hydrodynamics and gravitation. This study investigated the growth of the r-mode and found that the r-mode 
 amplitude reached values of order $1$  \citep{NOLINEVOLRMODES,NUMNONLINEVOL}. However, both Newtonian and relativistic 3-dim hydrodynamical simulations were (and still are) 
 severely limited by computational time and cannot probe the timescales on which r-modes may saturate. To bypass this limitation \text{Lindblom et al.\,\,\,}used 
 artificially enhanced gravitational radiation force (increased by a factor of 4500) in order to evolve the r-mode amplitude and allow it to reach values of order 1. \\  
 
 Later on the problem was attacked analytically with the use of a weakly non-linear perturbation theory to study the non-linear interactions of the 
 r-modes with other inertial modes \citep{NONLINCOUPLROTNS,SATRMODEINST,RMCOUPLENERGTRANS}. Calculations on a network of $\sim 5000$ modes with about 1.3 million couplings 
 showed that the r-mode amplitude saturated at small values of order $10^{-4}$. Furthermore, they showed that under some circumstances, 
 it is possible to show that the r-mode amplitude evolution is dominated by just one three-mode coupling (r-mode plus 2 daughter modes). \\
 
 The triplet of r-mode plus the 2 daughter modes, together with assumptions on the neutron star cooling mechanism and hyperon bulk viscosity, was used in further work on the r-mode instability: 
 (a) describing the rotational frequency evolution of accreeting neutron stars \citep{SDNS2007} and (b) model the spin-down of a newborn neutron star by the emission
 of r-mode gravitational radiation \citep{SDNS2009}. The work on (a) showed that accreeting neutron stars undergo a cyclic thermal runaway evolution during which 
 the r-mode amplitudes remain saturated at values of order $10^{-5}$ while work on (b) showed that the r-mode amplitudes in newborn neutron stars saturate at values of 
 order $10^{-4}$ to $10^{-2}$. \\ 
 
 Using a single triplet of modes, the lowest parametric instability threshold is very sensitive to the internal neutron star physics. The authors 
 considered a model with a neutrino cooling mechanism, via a combination of fast and slow processes, viscous heating due to hyperon bulk viscosity and spin-down due 
 to gravitational radiation and magnetic dipole radiation. Two necessary assumptions (that are also limitations) of this single-triplet model and were made so that 
 r-mode amplitude saturation is achieved are: (a) high hyperon bulk viscosity and (b) fast neutrino cooling mechanism. The fast neutrino cooling assumes certain 
 restrictions on the equation of state of the neutron star matter that may not be true for every neutron star. For example fast neutrino cooling by direct Urca-like 
 processes requires exotic phases of matter in the core of neutron stars \citep{NSC,NSCS}. Furthermore, if real neutron stars have low hyperon bulk viscosity 
 the energy dissipated by the single triplet of modes used in \citep{SDNS2007,SDNS2009} may not be sufficient to stop the growth of the r-mode amplitude. \\
 
 If the Bondarescu et \text{al.} assumptions for the neutron star model (based on the single triplet of modes) are not satisfied, several triplets of modes may be required to 
 saturate the r-mode amplitude. In that case, several parametric instability thresholds would be passed before saturating the r-mode amplitude thus making it not clear at what 
 order the latter would saturate. If the r-mode amplitude passes its first parametric instability threshold value, other near-resonant inertial modes are expected to be excited 
 (via energy transfer from the r-mode) signaling the importance of non-linear effects \citep{SDNS2009}. This is a fundamental property of inertial modes which does not depend on 
 the details of the equations of state or on a slow-rotation approximation. The r-mode frequencies reside inside an extended region where there is a sufficiently large number of 
 inertial modes with similar frequencies, so it is easy for resonance conditions to be satisfied. \\
 
 Even if several multiple triplets of modes are included in simulating the non-linear effects, depending on how reliable the mode-coupling estimates of the r-mode saturation amplitude is, 
 it may not be certain that the energy transfer to inertial modes will stop the instability at low r-mode amplitudes. If the mode-coupling mechanism does not saturate the growth 
 of the r-mode early enough \text{Bondarescu et al.\,\,\,}pointed out that other mechanisms like suprathermal bulk viscosity might then become relevant. The current scientific 
 consensus is that the mode-couplings saturating the amplitude of the r-mode is a fundamental result and will remain valid even with an improved understanding. Nevertheless, it 
 is thought very unlikely that the r-mode amplitude could grow larger than $10^{-1}$. \\ 
 
 In the absence of experimental evidence several authors still entertain the idea of r-modes saturating at amplitudes of order equal to or greater than $10^{-1}$. It was 
 shown that if the non-linear coupling mechanism does not come into play the r-mode amplitude would keep rising exponentially and eventually reach the suprathermal regime (where the chemical 
 equilibrium becomes greater than $K_B T$). In this high r-mode amplitude regime, the damping due to (suprathermal) bulk viscosity increases dramatically (non-linearly) with increasing 
 r-mode amplitude. This causes the viscous damping to overcome the gravitational instability and saturate the r-mode at amplitudes of order $10^{-1}-1$ \citep{ALFORD}. \\
 
 Considering the results found by the above studies it seems that the most likely values the r-mode amplitude may saturate at is anywhere 
 between $10^{-4}$ to $10^{-1}$. In our sensitivity study we considered r-mode waveforms of three saturation amplitudes $10^{-3}$, $10^{-2}$ and $10^{-1}$. Considering the 
 worst case scenarios (that r-mode amplitudes saturate at values less than $10^{-2}$) our best chances to settle this open question is to design a targeted search of 
 r-mode gravitational waves from newborn neutron stars. This requires some waiting time and some preparation in order to be ready for the next local group supernova. \\

\section{Gravitational wave detectors and simulated waveforms}

In this section we discuss the sensitivity level of the aLIGO and ET detectors   
We also present several plots of the frequency evolution dependence on $f_o$ and $\alpha$ (according to the \text{Owen et al.\,\,\,}'98 model)
and how these waveforms are used to construct the signals that were injected into the time-shifted data. 

\subsection{Sensitivity of the gravitational-wave interferometers}
The past decade has seen the development of a worldwide network of gravitational-wave interferometers.
The initial LIGO detectors achieved design strain sensitivity of $\unit[\sim2\times10^{-23}]{Hz^{-1/2}}$ in the most sensitive frequency range of 
$\unit[100-130]{Hz}$. The next generation of interferometers such as aLIGO and advanced Virgo (aVirgo) are expected to 
begin taking data starting in early Fall $2015$. The upgraded aLIGO/aVirgo experiments are expected to achieve a factor of ten higher strain sensitivity 
than initial LIGO/Virgo. In our simulation study below we show that aLIGO can probe r-mode gravitational-wave spin-down signals to newborn neutron stars within the 
Local Group of galaxies. \\

\begin{figure}[hbtp!]
\epsscale{1.3}
\plotone{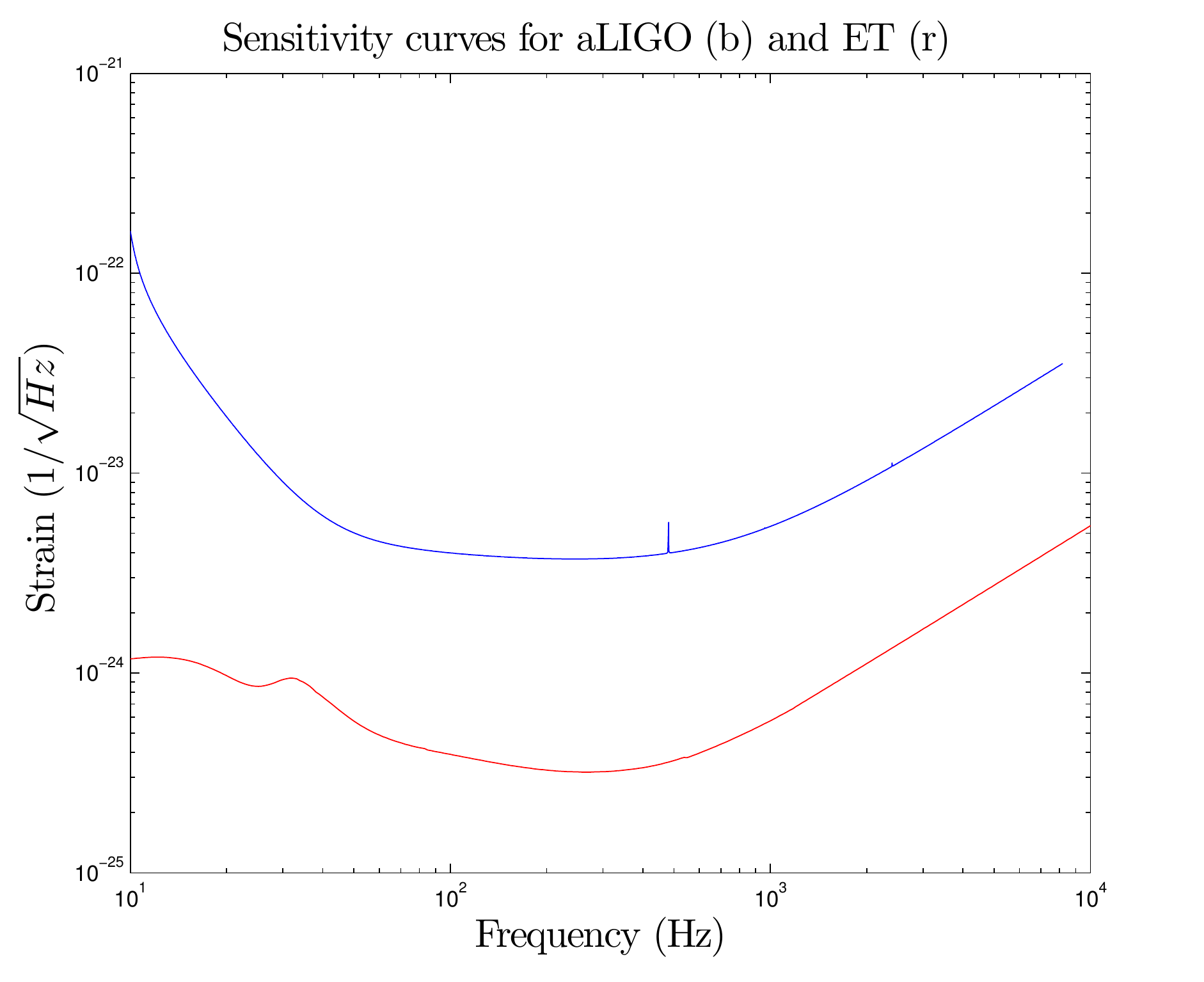}
\caption{The sensitivity curves of aLIGO and ET: aLIGO simulated noise shows a maximum sensitivity of $\unit[\sim5\times10^{-24}]{Hz^{-1/2}}$ at $\unit[110-130]{Hz}$
while ET simulated noise shows a maximum sensitivity of $\unit[\sim 2\times10^{-25}]{Hz^{-1/2}}$ at $\unit[110-130]{Hz}$.} \label{Fig:ef3}
\epsscale{1.2}
\end{figure}

\subsection{Choice of parameter values and waveform simulation}

The \text{Owen et al.\,\,\,}'98 model was used to produce the waveforms injected in the MC and eLIGO noise colored with the aLIGO sensitivity 
curve and also the eLIGO noise colored with the ET sensitivity curve. In this model there are two parameters: the initial frequency $f_o$ and 
the saturation amplitude $\alpha$. We assume that the gravitational-wave frequency of the signal is between $\unit[600-1600]{Hz}$ (as shown in 
table \ref{tab:distances}). We also assume that the signal will be powerful enough for detection for at least $\unit[2500]{s}$. This signal 
duration is enough for the power to drop to $\unit[17]{\%}$ of its initial value for waveforms with highest $f_o$ and $\alpha$ values. 
However, the power drops only to $\unit[87]{\%}$ of the initial value for waveforms with the lowest $f_o$ and $\alpha$ values. In the latter case, 
longer durations would be required to increase the SNR of the signal (when clustering pixels in the detection algorithms) and hence increase the 
detection distance. However, using $\unit[2500]{s}$ duration and $\unit[600-1600]{Hz}$ frequency span results in ft-maps of $2.5\times10^6$ pixels. 
To run the background and sensitivity study algorithms and to create ft-maps of this size requires memory that approaches the memory limitations 
of conventional CPUs. \\

\begin{figure}[hbtp!]
\epsscale{1.3}
\plotone{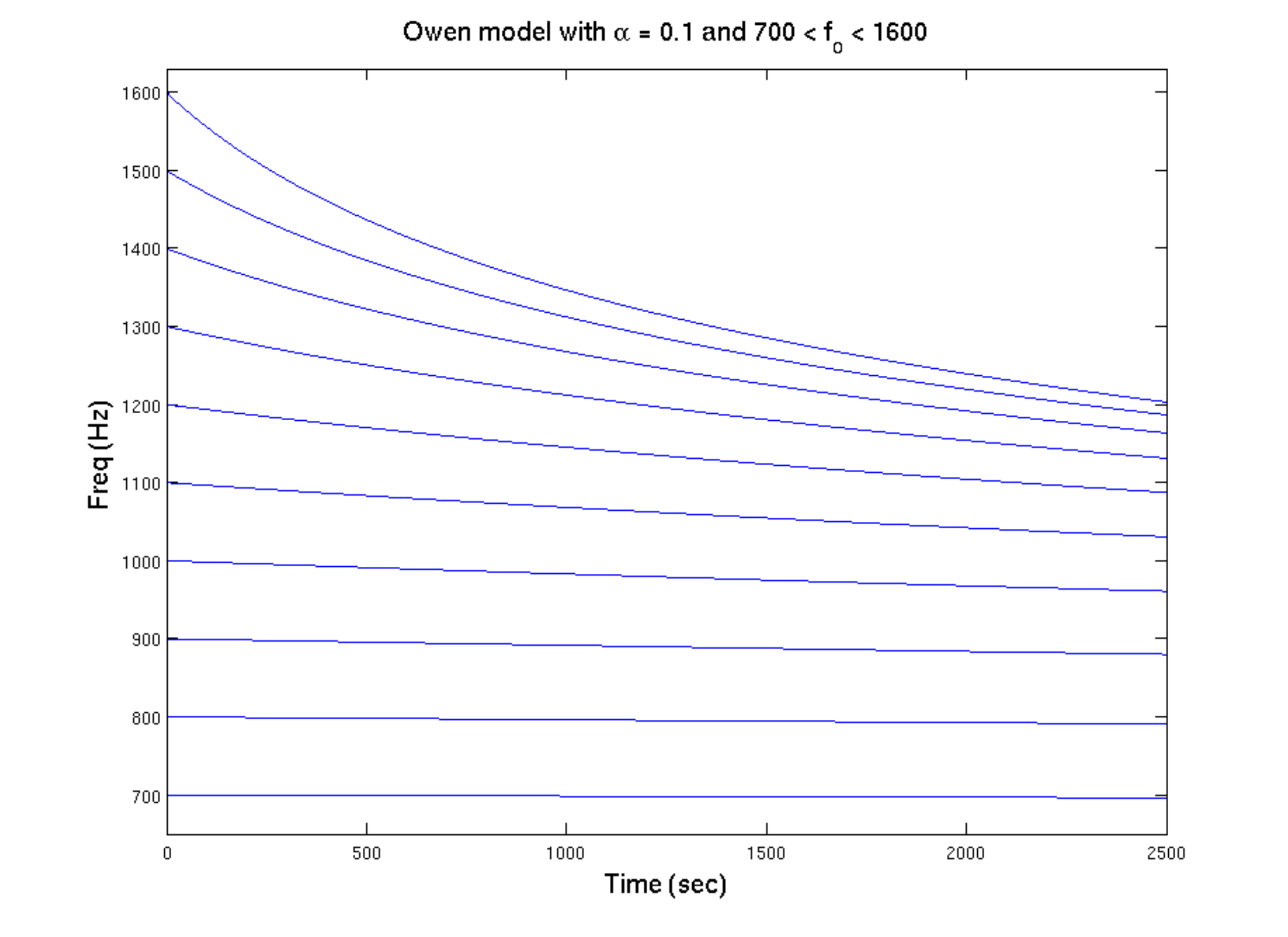}
\caption{Frequency evolution dependence on the initial frequency $f_o$. The higher the $f_o$ value the steeper the spin-down.} \label{Fig:ef4}
\epsscale{1.2}
\end{figure}

\begin{figure}[hbtp!]
\epsscale{1.3}
\plotone{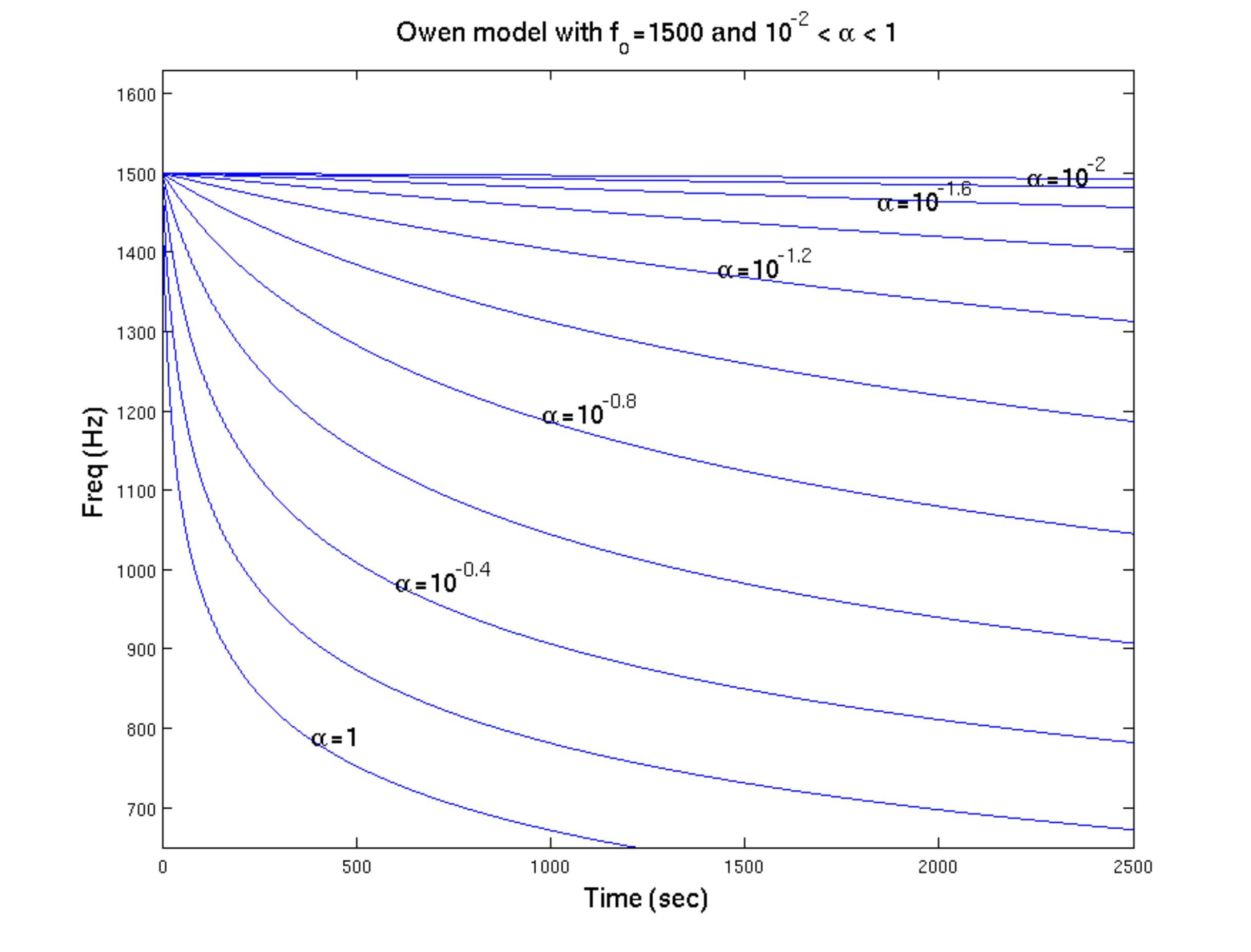}
\caption{Frequency evolution dependence on the saturation amplitude $\alpha$. The higher the $\alpha$ value the steeper the spin-down.} \label{Fig:ef5}
\epsscale{1.2}
\end{figure}

We used three $f_o$ values 
(700, 1100, 1500 Hz) and three $\alpha$ values ($10^{-3}$, $10^{-2}$, $10^{-1}$) resulting in nine waveforms. The dependence of the waveforms on $f_o$ and $\alpha$ are shown 
in Fig.\ref{Fig:ef4} and Fig.\ref{Fig:ef5}

A gravitational wave emitted by a neutron star is a linear combination of the two polarizations, $h_+$ and $h_{\times}$ and can be expanded in the detector's
frame like \citep{OWENBB}

\begin{equation}
\label{eq.23}
h(t) = F_+ (t, \psi) h_+ (t) + F_{\times} (t, \psi) h_{\times} (t)
\end{equation}

\noindent
where the coefficients $F_+ (t, \psi)$ and $ F_{\times} (t, \psi) $ are the detector antenna pattern functions for the two polarizations, t is the time 
in the detector frame and $\psi$ is the polarization angle. The waveforms for the two polarizations are given by

\begin{equation}
\label{eq.24}
 h_+(t) = h_o \left ( \frac{1+ \cos^2\theta}{2} \right ) \cos \Phi(t) 
\end{equation}

\noindent
and

\begin{equation}
\label{eq.25}
 h_{\times}(t) = h_o \cos(\theta) \sin\Phi(t)
\end{equation}

\noindent
where $h_o$ is the corresponding gravitational-wave strain amplitude as measured on Earth, $\theta$ is the inclination angle and $\Phi(t)$ is the 
time dependent phase of the gravitational wave in the detector's frame and is given by

\begin{equation}
\label{eq.26}
 \Phi(t)= \Phi (t_0) + \Delta \Phi = \Phi(t_0) + \int_{t_o}^t 2 \pi f_{gw}(t') dt'
\end{equation}

\noindent
where $t_0$ is the fiducial start time of the observation and $f_{gw}$ is the gravitational wave frequency in the detector's frame. The angle $\theta$ can be set 
equal to zero by assuming that the direction of propagation of the gravitational wave is perpendicular to the plane of the interferometer. \\

\begin{figure}[hbtp!]
\epsscale{1.3}
\plotone{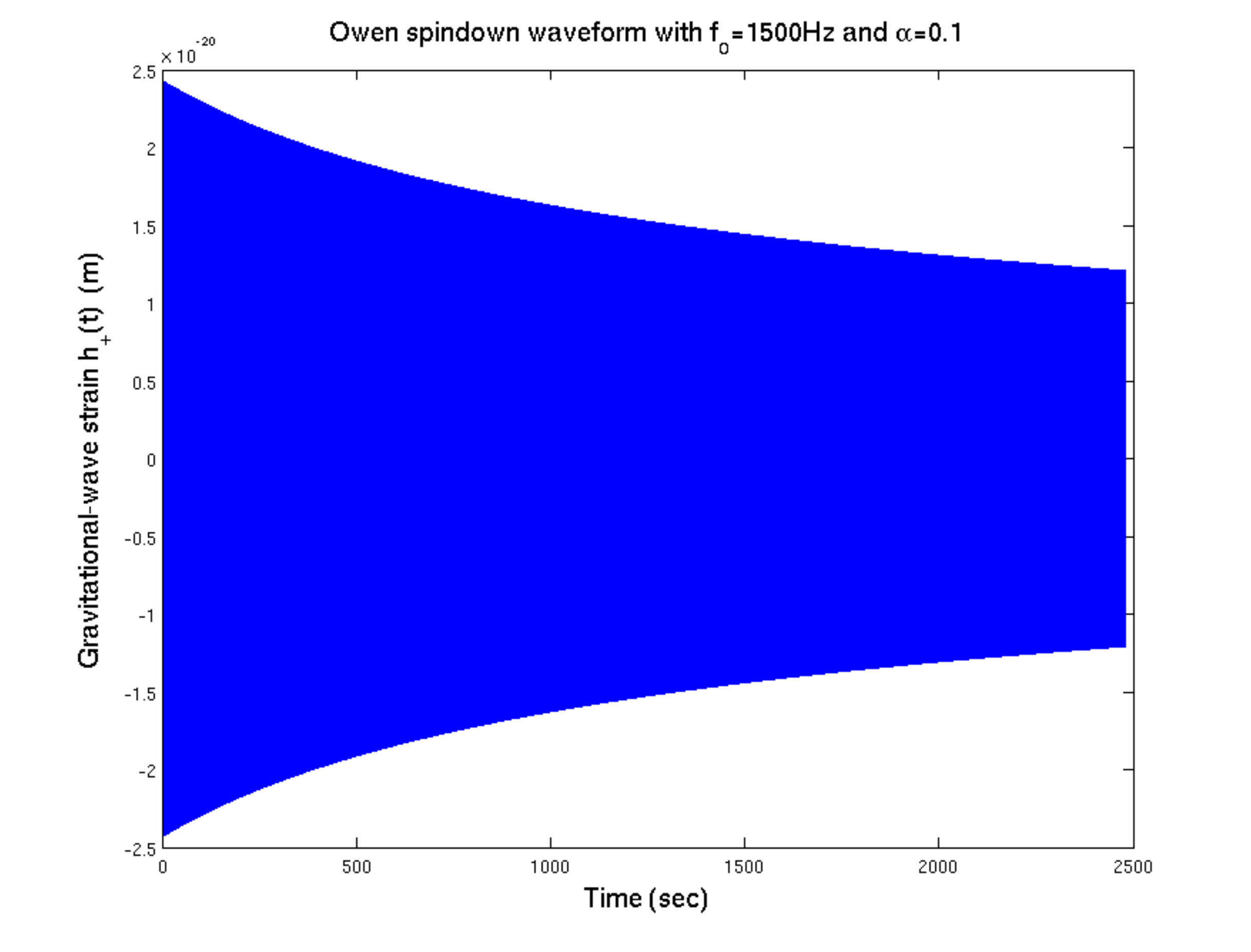}
\caption{This is an example of a $h_{+}(t)$ gravitational-wave strain that was injected in simulated aLIGO and ET noise.} \label{Fig:ef6}
\epsscale{1.2}
\end{figure}

\section{Injection recovery and detectability}

This section starts with a discussion on matched filtering, the optimal detection algorithm. This algorithm requires accurate knowledge of the signal we 
are searching for. Though this is not the case for the r-modes (due to the wide range of possibilities for the parameter values) matched filtering gives 
theoretical upper bounds and a measure of evaluating the efficiency of the clustering algorithm used in our present sensitivity study. 
Our study is designed to assess the detectability of r-mode gravitational wave signals injected on (time-shifted) data colored with the sensitivity curves
of aLIGO and ET gravitational wave detectors. Focusing on realistic excess cross-power searches, we show that second-generation detectors such as 
advanced LIGO (aLIGO) may be able to probe neutron star spin-down models out to astrophysically interesting distances of $\unit[\sim1]{Mpc}$. Third generation 
detectors such as Einstein Telescope (ET) may be able to probe such signals out to distances of $\unit[\sim10]{Mpc}$. \\

\subsection{Matched filtering estimates}
For gravitational-wave signals of a known form, the theoretically optimal search strategy is matched filtering \citep{MF}. However, the r-mode signals we consider 
here are not suitable for a search with a matched filtering detection algorithm due to the lack of accurate models that are available to generate r-mode waveforms. 
A search with matched filtering must include a template bank that spans the space of all possible signals and unknown parameter values. An incomplete template bank 
can result in faulty upper limits if the true signal falls outside the template space. In order to generate a complete template bank, we require firm knowledge of 
the details of the waveform's phase evolution. In presenting the above model, we have not aspired to this degree of accuracy. Moreover, even if a complete and 
accurate model could be written down, there may be computational challenges associated with performing the search, especially for long signals with many parameters. \\

Nonetheless, it is useful to compare the detection distances calculated using the excess-power technique to estimates for what can be achieved with matched filtering 
as this places an upper limit on the detection distance that can be achieved through improvements to the data-analysis scheme. As before, we assume optimal orientation 
of the source, optimal orientation of the detector network, a false-alarm probability $ \leq \unit[0.1]{\%}$ and a false dismissal probability $ \leq \unit[50]{\%}$. 
We find that highly idealized matched filtering allows us to extend the detection distance up to factors of 10 - 20 as shown in table 1. This factor varies depending 
in part on the efficiency of the pattern recognition algorithm for different signal types. \\

\subsection{Spectrograms and the seedless clustering detection algorithm }

To estimate the distances at which we can see the signals predicted by the \text{Owen et al.\,\,\,}'98 model using various combinations of the model parameters 
$(f_o,\alpha)$, we simulate an excess cross-power search for long-lived gravitational waves associated with a well-localized electromagnetic counterpart, 
in the r-mode case this would be a supernova explosion. This simplifies the search to a single direction, although these techniques may be extended to 
all-sky searches in the future. \\

\begin{figure}[hbtp!]
\epsscale{1.2}
\plotone{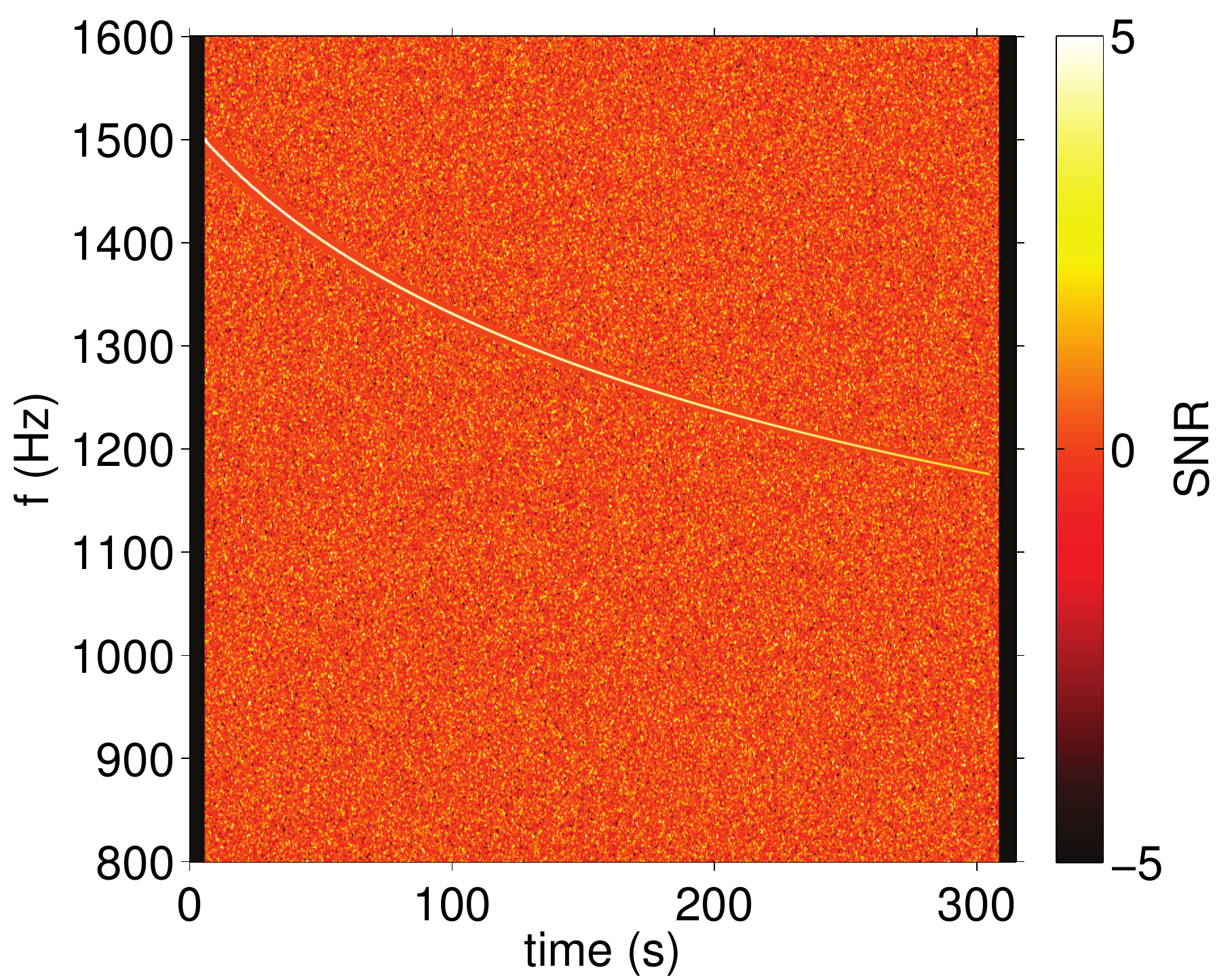}
\caption{Injection of a r-mode signal using a waveform of $f_o=\unit[1500]{Hz}$ and $\alpha=0.1$ at a distance of $\unit[1]{Kpc}$.
The background was simulated using MC data colored with the aLIGO sensitivity curve.} \label{Fig:ef7}
\epsscale{1.2}
\end{figure}

Within this $ \unit[2500]{s} \times \unit[1000]{Hz}$ on-source region, and following \citep{STAMPPAPER}, we create a spectrogram of signal-to-noise ratio $\text{SNR}(t;f)$,
which is proportional to the cross-correlation of the H1 and L1 strain data. Here $f$ refers to the frequency bin in a discrete Fourier transform centered on time $t$. 
We use $\unit[1]{s}$-long, $\unit[50]{\%}$-overlapping, Hann-windowed data segments, which yield spectrograms with a resolution of $ \unit[0.5]{s} \times \unit[1]{Hz}$. 
The signal to noise ratio is given by

\begin{equation}
  \langle\text{SNR}(t;f)\rangle = h_0^2 \left( 1-\cos^2\iota \right) ,
\end{equation}

\noindent
 where $\iota$ is the angle between the direction of propagation of the gravitational wave and the plane of the interferometer (the noise fluctuations can be both positive 
 and negative). A formal derivation of $\text{SNR}(t;f)$ can be found in \citep{STAMPPAPER}. An example of a SNR(t;f) spectrogram 
 with a r-mode injection added to simulated detector noise can be seen in Fig.\ref{Fig:ef7}. The problem of detection is to identify a track of positive-valued pixels in
 the presence of noise. \\

 In our simulation, we use a network consisting of two 4km LIGO observatories assumed to be operating at aLIGO design sensitivity; one in Hanford (H1) and one in 
 Livingston (L1). We create a spectrogram of signal-to-noise ratio SNR(t;f) in our $\unit[2500]{s} \times \unit[1000]{Hz}$ on-source region, which is proportional to the cross-correlation
 of the H1 and L1 strain data. \\

 To recover the injected signals, we apply a seedless clustering algorithm \citep{stochtrack} that looks for clusters of positive pixels. 
 The pixels are combined to determine the SNR for the entire cluster denoted by $\text{SNR}(c)$ which is distinct from $\text{SNR}(t;f)$ associated with individual 
 pixels in a spectrogram. To form the SNR for the cluster the pixels are summed using their individual sigmas as weights. \\
 
 The algorithm was applied to Monte Carlo Gaussian noise colored with the aLIGO sensitivity curve, eLIGO S5 data (noise) recolored with the aLIGO sensitivity curve
 and Monte Carlo Gaussian noise colored with the ET sensitivity curve (Fig.\ref{Fig:ef3}). The algorithm was used in order
 to determine the threshold for an event with false alarm probability $ \leq \unit[0.1]{\%}$. For the MC data colored with the aLIGO sensitivity curve we found 
 the cluster SNR threshold to be $7.2$ while using the ET sensitivity curve we found the cluster SNR threshold to be $8.1$. \\

We then injected simulated signals into the Monte Carlo noise at varying distances. We determined the distance above which $\unit[50]{\%}$ of the signals are recovered, 
which corresponds to the distance at which we can observe a signal with false alarm probability $ \leq \unit[0.1]{\%}$ and false dismissal probability $ \leq \unit[50]{\%}$. 
These detection distances are shown in table 1 for all 9 different combinations of parameters. \\

\begin{deluxetable*}{c c c c c}
\tablecolumns{5} \tablewidth{0pt}
\tablecaption{Sensitivity study using a seedless clustering (SC) algorithm and matched filtering (MF)}
\tablehead{
\colhead{r-modes waveform}      & \colhead{MC data}          & \colhead{eLIGO data}    &   \colhead{MC data}  & \colhead{MC data} \\
\colhead{($f_{o}$, $\alpha$)}   & \colhead{aLIGO s.c./SC}       & \colhead{aLIGO s.c./SC}    &   \colhead{ET s.c./SC}  & \colhead{aLIGO s.c./MF}  \\
\colhead{(Hz, unitless)}        & \colhead{(Mpc)}            & \colhead{(Mpc)}         &   \colhead{(Mpc)}    & \colhead{(Mpc)}  \\
 }  
\startdata

(1500, 0.1)   &  1.2    &  1.1   & 9.7  &  22   \\       
(1100, 0.1)   &  0.97   &  1.1   & 8.1  &  13   \\
(700, 0.1)    &  0.44   &  0.42  & 4.3  &  4.8  \\
 
(1500, 0.01)  &   0.19  &  0.21  &  1.8  &  2.9    \\
(1100, 0.01)  &   0.13  &  0.12  &  1.1  &  1.3    \\   
(700, 0.01)   &   0.040  &  0.043 &  0.39 &   0.54  \\       
                                         
(1500, 0.001)  &   0.016  &  0.021  &   0.16   &  0.35   \\
(1100, 0.001)  &   0.014  &  0.012  &   0.11  &  0.19   \\
(700, 0.001)   &   0.0040  &  0.0044  &   0.039  &  0.048

  \enddata
   \tablenotetext{a}{Distances are calculated for a false alarm probability $ \leq \unit[0.1]{\%}$ and a false dismissal probability $ \leq \unit[50]{\%}$ using Monte Carlo (MC) aLIGO noise and MC ET noise.} 
   \tablenotetext{b}{Doing a background study on 1000 maps the cluster SNR threshold was estimated to be $SNR_{th} = 7.2$ with the aLIGO sensitivity curve (s.c.) and $SNR_{th} = 8.1$ 
   with the ET s.c. Sources are assumed to be optimally oriented.}
   \tablenotetext{c}{Columns 2, 3 and 4 are the distances at which 50 out of 100 injections were recovered using the SC algorithm and column 5 shows the corresponding distances using a highly idealized MF.}
   
   \label{tab:distances}
\end{deluxetable*}

When testing the sensitivity of a given detection algorithm, we run the algorithm on $10^3$ noise maps and out of those we set the highest (cluster) SNR as the SNR threshold.
This threshold corresponds to a fixed gravitational-wave strain, $h_o=h_{th}$. Using \eqref{eq.15} and this threshold $h_{th}$, we can test what combination of
parameters $\alpha$ and $f_o$ and at what distance d the corresponding waveform can be detected. Substituting $h_o$ with $h_{th}$ and solving \eqref{eq.15}
for $d$ we get the distance expression in units of Mpc

\begin{equation}
\label{eq.27}
 d \approx 1.5 \times 10^{-23} \left( \frac{f}{1kHz}  \right )^3 \frac{| \alpha |} {h_{th}}
\end{equation}

\noindent
where

\begin{equation}
\label{eq.28}
  f^3 \sim ( 1 + 10^{-20} |\alpha|^2 f_o^{6}t ) ^{ - \frac{1}{2} } 
\end{equation}

Equations \eqref{eq.27} and \eqref{eq.28} suggest that (for a given value of $h_{th}$) knowing the detection distance $d_1$ for one waveform of parameters $(\alpha_1, f_{o1})$
we can estimate the expected distance for another waveform of parameters $(\alpha_2, f_{o2})$ using

\begin{equation}
\label{eq.29}
d_2 \sim \frac{|\alpha_2|}{|\alpha_1|} \left ( \frac{1+ \lambda |\alpha_1|^2 f_{o1}^6}{1+ \lambda |\alpha_2|^2 f_{o2}^6} \right)^{\frac{1}{2}} d_1
\end{equation}

\noindent
where $\lambda \approx 10^{-20} t$. This suggests that when the parameters $(\alpha, f_{o})$ are such that $ \lambda |\alpha|^2 f_{o}^6 \ll 1 $ then 

\begin{equation}
\label{eq.30}
d_2 \sim \frac{|\alpha_2|}{|\alpha_1|} d_1
\end{equation}

\noindent
This is the case for $f_{o1}=7\times10^2$, $\alpha_1=10^{-3}$ and $t=2.5\times10^3$ giving $ \lambda |\alpha_1|^2 f_{o1}^6 \approx 2.9 \times 10^{-6} $ and $f_{o2}=7\times10^2$, $\alpha_2=10^{-2}$ and $t=2.5\times10^3$ 
giving $ \lambda |\alpha_1|^2 f_{o1}^6 \approx 2.9 \times 10^{-4} $. The seedless based algorithm gives for the first case a detection distance of $\unit[40]{Kpc}$ while for the second case
it gives $\unit[4]{Kpc}$ according to \eqref{eq.30}. \\

For $ \lambda |\alpha|^2 f_{o}^6 \simeq 1 $ or $ \lambda |\alpha|^2 f_{o}^6 \ge 1 $ the detection distances our algorithm gives follow equation \eqref{eq.29}. For example 
for $f_{o1}=1.5\times10^3$, $\alpha_1=10^{-1}$ and $t=2.5\times10^3$ giving $ \lambda |\alpha_1|^2 f_{o1}^6 \approx 2.9 $ and $f_{o2}=1.5\times10^3$, $\alpha_2=10^{-2}$ and $t=2.5\times10^3$ 
giving $ \lambda |\alpha_1|^2 f_{o1}^6 \approx 3\times10^{-2} $ we get the relation $ d_2^{est} \approx 0.19 d_1$. The detection distances $d_1$ and 
$d_2$ are related by $ d_2 \approx 0.17 d_1$ (column 2), $ d_2 \approx 0.19 d_1$ (column 3) and $ d_2 \approx 0.18 d_1$ (column 4) of table 1. Thus we conclude that the 
detection algorithm behaves in a way consistent with the theoretical predictions. \\

\section{Discussion}
\label{sec:end}

The most optimistic scenarios in the \text{Bondarescu et al.\,\,} numerical simulations show that r-mode gravitational-wave strains take values of $\unit[5 \times10^{-25}]{Hz^{-1/2}}$ at $f=\unit[400]{Hz}$ 
and $\unit[6 \times10^{-24}]{Hz^{-1/2}}$ at $f=\unit[600]{Hz}$ from sources that are $\unit[100]{Kpc}$ away from the detectors. Using \eqref{eq.15} we can show that these strain amplitudes correspond to r-mode saturation 
amplitudes that range from $\alpha = 5.2 \times 10^{-2}$ to $\alpha = 1.9 \times 10^{-1}$. In the worst case scenarios their work shows that r-mode gravitational wave strains take 
values of $\unit[5 \times10^{-26}]{Hz^{-1/2}}$ at $f=\unit[600]{Hz}$ ($\alpha = 1.5 \times 10^{-3}$) to $\unit[2\times10^{-27}]{Hz^{-1/2}}$ at $f=\unit[350]{Hz}$ ($\alpha = 3.1 \times 10^{-4}$). Clearly, our results show that 
we can only be hopeful for a large $\alpha$ r-mode gravitational wave detection. \\

In the non-linear bulk viscosity mechanism of \text{Mark Alford et al.\,\,}the r-mode amplitudes grow up to values of orders $10^{-1}$ - $1$. A $\unit[1]{KHz}$ gravitational-wave signal 
generated by r-mode oscillations saturated at these amplitudes will create gravitational-wave strain values from  $10^{-24}$ to $10^{-23}$ respectively at distances of $\unit[1]{Mpc}$ 
where the supernova rate is about 3-4 per century. The same wave at a distance of $\unit[10]{Mpc}$ where the supernova rate is at about 1-2 per year \citep{SNRATE} will have a strain 
value of order $10^{-25}$ to $10^{-24}$. Our sensitivity results show that we can exclude the O(1) r-mode oscillations by the absence of any r-mode detections at those distances.\\

Our results show that for a $\unit[1]{KHz}$ gravitational wave generated by a r-mode oscillation saturated at an amplitude of 
order $10^{-2}$ the maximum detection distance is $\unit[0.1]{Mpc}$. The result for a gravitational wave of the same frequency but generated by a r-mode oscillation saturated 
at an amplitude of order $10^{-1}$ is a maximum detection distance of $\unit[1]{Mpc}$. These results show that our search method can detect gravitational-wave strains of order down to $10^{-24}$. 
The sensitivity curve of aLIGO at $\unit[1]{KHz}$ is about $\unit[5 \times10^{-24}]{Hz^{-1/2}}$ showing that with our current algorithms we can detect signals that are 5 times weaker than the noise. \\

Detecting a gravitational-wave strain of order $10^{-24}$ gives information about the ratio $\alpha / d$ and does not specify the value of $\alpha$. Therefore, 
to determine the value of $\alpha$ we need an electromagnetic trigger to provide the distance for us. Given the distance we can then calculate the value of the 
saturation parameter, $\alpha$. A detection of r-mode gravitational radiation coming from a supernova is the best possible outcome of our efforts in this search. 
However, a non-detection can also be useful in determining an upper limit on the r-mode saturation amplitude and disprove the theories that suggest scenarios 
of r-mode saturation amplitudes higher than that. \\

Initially, this project started with the target to design a search that would set an upper bound on the saturation amplitude of the r-mode mass current oscillations on 
neutron stars. For this work we used the Owen et \text{al.\,\,}'98 model that assumes a polytropic equation of state (EOS). Therefore, our work was initially based on a specific EOS, using 
specific values for $\tilde{J}_2$ and $\tilde{I}$ giving $Q \sim 0.094$. However, we may use different EOS and using the integrals for $\tilde{J}_2$ and $\tilde{I}$ we can 
calculate their corresponding $Q$ value and therefore, the corresponding waveforms. Therefore, a hypothetical r-mode gravitational wave detection can not only settle the 
debate on the r-mode oscillations saturation amplitude but also impose severe constraints on the EOS of the matter in the core of a neutron star. \\

In the Owen et \text{al.\,\,}'98 model, non-linear couplings between the r-mode and other inertial modes 
as well as strong magnetic fields (crucial in the evolution/saturation of the r-mode amplitude) were not explicitly modeled. However, when compared to 
the complicated numerical simulations in Bondarescu et al. we got a significant overlap for the r-mode gravitational wave frequency evolution during the first two weeks after 
the neutron star is born. This is the era of a neutron star's life we are interested in. \\

An improved model that includes magnetic fields effects can be found in \citep{HKTOP}. This work considers the same Owen et \text{al.\,\,}'98 model but with an additional frequency 
damping term due to magnetic braking. The restrictions their findings imply are that the r-modes may be dominant only for B-fields of magnitude below $\unit[10^{13}]{G}$. This 
result does not impose any changes on the r-mode gravitational waveforms. It merely says that we shouldn't be hopeful to observe r-mode gravitational radiation in neutron  
stars with very strong magnetic fields. On the other hand, this magnetic braking restriction is dependent on the equation of state, hence not very decisive on the magnitude 
of the magnetic fields for which we should not hope to detect r-mode gravitational waves. \\ 

The accuracy of the waveforms determines the type of decision making algorithm we should be using when searching for a signal. Lack of reliable models of the physical system
implies production of unreliable waveforms. This is the reason we did not search for the r-mode signal using matched filtering. The algorithms we used in this study are 
based on the statistical significance of signal to noise ratios of clusters made of pixels above a certain snr threshold. This method did not use any knowledge of the signal. 
Knowledge of the r-mode signal can be used and make minor modifications in our existing algorithms, however, there was not much hope for a dramatic improvement in
sensitivity. We were still able to recover signals of one order of magnitude weaker than the noise. We are currently considering algorithms designed to search for specific 
signals, in our case the r-mode. We hope that customizing the search algorithm for a specific signal can significantly increase the sensitivity in decision making. \\

An ideal class of decision making algorithms, which are suitable specifically for cases when the signal is not precisely (but only crudely) known, is provided by machine 
learning algorithms (MLAs). Three classes of MLAs we are currently exploring are artificial neural networks (ANN), support vector machines (SVM) and local space classifiers (LSC). 
MLAs are cosidered novel methods of signal recognition in the area of gravitational-wave searches. We are now working on developing r-mode search algorithms using the 
above three techniques. Currently, we are producing data in the parameter space of interest. The plan is to repeat the same sensitivity study we presented in this paper 
using MLAs for decision making/recognition of the desired signals. The results will be compared to our present results and published in a future paper. \\

\section{Conclusions}

(a) In the worst case scenario of \text{Bondarescu et al.\,\,}, $\alpha =10^{-4} - 10^{-3}$. If this is correct the results from our sensitivity study show that we can only be 
hopeful (with this search method) to detect r-mode gravitational radiation from sources that are up to $\unit[21]{Kpc}$ away. \\

(b) If the best case scenario of \text{Bondarescu et al.\,\,}(with $\alpha=10^{-1}$) prevails then we can be hopeful to detect r-mode gravitational radiation from sources that are up to $\unit[1100]{Kpc}$ away.
In this case it may not be possible to distinguish whether it is \text{Bondarescu et al.\,\,}or \text{Alford et al.\,\,}mechanism that comes into play. \\

(c) To exclude the validity of \text{Bondarescu et al.\,\,}results a r-mode gravitational wave detection yielding a value of $\alpha \ge 0.2$ is needed. In that case we will have strong indications that 
\text{Alford et al.\,\,}mechanism is the correct r-mode oscillation amplitude saturation mechanism. \\

(d) If the best case scenario (of high r-mode saturation amplitudes) is correct we may be able to detect r-mode gravitational radiation from a supernova event within our local group ($\unit[\sim1]{Mpc}$). 
Considering the low rate of the local group supernovae (3-4 per century) and that the latest supernova in the local group occurred in January 2014, we may have to wait another 25-30 years until the next one. \\

(e) The aLIGO and ET sensitivity results show that for large amplitudes of $\alpha$, the detection distances are up to $\unit[1.1]{Mpc}$ (2-3 events per century) and $\unit[10]{Mpc}$ (1-2 events per year) respectively. 
Using aLIGO, to exclude saturation amplitudes of order $10^{-1}$, a null result at a distance of $\le \unit[0.42]{Mpc}$ is required. Using ET, to exclude the same saturation amplitudes a null result at a distance of 
$\le \unit[4.3]{Mpc}$ is required. Considering the event rates at those distances we can only hope for such results only after several years of ET operation. \\

(f) We do not need all the complicated physics of the neutron star to approximate the r-mode waveforms during the first 2 weeks after the onset of the r-mode gravitational radiation. A model where high magnetic fields
are excluded and details of the r-mode saturation mechanism are unknown can still give a good approximation of the r-mode waveforms during the early stages of gravitational radiation. \\

(g) When the model waveform parameters take values over large ranges, choosing a decision making algorithm may not be a simple task. Clustering algorithms may yield satisfactory results but that is not necessarily
the best option we can have. Other techniques (for example MLAs) that do not require accurate knowledge of the waveform parameter values may result in longer detection distances. Increasing the detection distance by 
a factor of 2 or 3 can make the difference. \\

For lengthy discussions we thank: the Stochastic Transient Analysis Multi-detector Pipeline (STAMP) group that also developed the code we used, Ruxandra Bondarescu, 
Ira Wasserman, Mark Alford, Lee Lindblom, Kostas Kokkotas and Nick Stergioulas. \\


\begin{thebibliography}{}
\expandafter\ifx\csname natexlab\endcsname\relax\def\natexlab#1{#1}\fi

\bibitem[{Alford {et~al.}(2012)Alford, Mahmoodifar, \& Schwenzer}]{ALFORD}
Alford, M.~G., Mahmoodifar, S., \& Schwenzer, K. 2012, Phys.Rev., D85, 044051

\bibitem[{Alford \& Schwenzer(2012)}]{ALFORD13}
Alford, M.~G., \& Schwenzer, K. 2012, PoS, ConfinementX, 258

\bibitem[{Andersson(1998)}]{NCRM1997}
Andersson, N. 1998, Astrophys.J., 502, 708

\bibitem[{Andersson \& Kokkotas(2001)}]{RMODEINST}
Andersson, N., \& Kokkotas, K.~D. 2001, International Journal of Modern Physics
  D, 10, 381

\bibitem[{Ando {et~al.}(2005)Ando, Beacom, \& Yuksel}]{allrates}
Ando, S., Beacom, J.~F., \& Yuksel, H. 2005, Phys.Rev.Lett., 95, 171101

\bibitem[{Arras {et~al.}(2003)Arras, Flanagan, Morsink, Schenk, Teukolsky,
  {et~al.}}]{SATRMODEINST}
Arras, P., Flanagan, E.~E., Morsink, S.~M., {et~al.} 2003, Astrophys.J., 591,
  1129

\bibitem[{Baumgarte {et~al.}(1996)Baumgarte, Teukolsky, Shapiro, Janka, \&
  Keil}]{SHAPIRO}
Baumgarte, T., Teukolsky, S., Shapiro, S., Janka, H., \& Keil, W. 1996,
  Astrophys.J., 468, 823

\bibitem[{Bondarescu {et~al.}(2007)Bondarescu, Teukolsky, \&
  Wasserman}]{SDNS2007}
Bondarescu, R., Teukolsky, S.~A., \& Wasserman, I. 2007, Phys.Rev., D76, 064019

\bibitem[{Bondarescu {et~al.}(2009)Bondarescu, Teukolsky, \&
  Wasserman}]{SDNS2009}
---. 2009, Phys.Rev., D79, 104003

\bibitem[{Branch \& Tammann(1992)}]{TYPE1DIST}
Branch, D., \& Tammann, G.~A. 1992, Annual Review of Astronomy and
  Astrophysics, 30, 359

\bibitem[{Brink {et~al.}(2004)Brink, Teukolsky, \&
  Wasserman}]{RMCOUPLENERGTRANS}
Brink, J., Teukolsky, S.~A., \& Wasserman, I. 2004, Phys.Rev., D70, 121501

\bibitem[{Chandrasekhar(1970)}]{SOLTWOPRBL}
Chandrasekhar, S. 1970, Phys. Rev. Lett., 24, 611

\bibitem[{Clemens \& Rosen(2004)}]{ROSEN}
Clemens, J.~C., \& Rosen, R. 2004, Astrophys.J., 609, 340

\bibitem[{de~Araujo {et~al.}(2005)de~Araujo, Miranda, \& Aguiar}]{FMODE}
de~Araujo, J. C.~N., Miranda, O.~D., \& Aguiar, O.~D. 2005, Class.Quant.Grav.,
  22, S471

\bibitem[{Dimmelmeier {et~al.}(2008)Dimmelmeier, Ott, Marek, \&
  Janka}]{DimmOtt}
Dimmelmeier, H., Ott, C.~D., Marek, A., \& Janka, H.-T. 2008, Phys.Rev., D78,
  064056

\bibitem[{Dragicevich {et~al.}(1999)Dragicevich, Blair, \& Burman}]{galacticR}
Dragicevich, P.~M., Blair, D.~G., \& Burman, R.~R. 1999, Mon. Not. Roy. Astron.
  Soc., 302, 693

\bibitem[{Duncan(1998)}]{DUNCAN}
Duncan, R.~C. 1998, arXiv:astro-ph/9803060

\bibitem[{Ferrari {et~al.}(2003)Ferrari, Miniutti, \& Pons}]{FERRARI}
Ferrari, V., Miniutti, G., \& Pons, J.~A. 2003, Mon.Not.Roy.Astron.Soc., 342,
  629

\bibitem[{Friedman \& Morsink(1998)}]{JFSM1998}
Friedman, J.~L., \& Morsink, S.~M. 1998, Astrophys.J., 502, 714

\bibitem[{{Friedman} \& {Schutz}(1978{\natexlab{a}})}]{LAGRANGEPERT}
{Friedman}, J.~L., \& {Schutz}, B.~F. 1978{\natexlab{a}}, \apjl, 221, L99

\bibitem[{{Friedman} \& {Schutz}(1978{\natexlab{b}})}]{SECINSTROTNS}
---. 1978{\natexlab{b}}, \apj, 222, 281

\bibitem[{{Hashimoto} {et~al.}(1994){Hashimoto}, {Oyamatsu}, \&
  {Eriguchi}}]{UPLANGV}
{Hashimoto}, M.-A., {Oyamatsu}, K., \& {Eriguchi}, Y. 1994, \apj, 436, 257

\bibitem[{Heger {et~al.}(2000)Heger, Langer, \& Woosley}]{Hegetal2}
Heger, A., Langer, N., \& Woosley, S. 2000, Astrophys.J., 528, 368

\bibitem[{Heger {et~al.}(2003)Heger, Woosley, Langer, \& Spruit}]{HegEtal}
Heger, A., Woosley, S., Langer, N., \& Spruit, H. 2003, arXiv:astro-ph/0301374

\bibitem[{Ho \& Lai(2000)}]{HOLAI}
Ho, W.~C., \& Lai, D. 2000, The Astrophysical Journal, 543, 386

\bibitem[{{Kirshner} \& {Kwan}(1974)}]{photospheric}
{Kirshner}, R.~P., \& {Kwan}, J. 1974, \apj, 193, 27

\bibitem[{Kokkotas \& Andersson(2001)}]{OSCINSTRELST}
Kokkotas, K.~D., \& Andersson, N. 2001, arXiv:gr-qc/0109054

\bibitem[{Lattimer \& Prakash(2001)}]{LatPrak}
Lattimer, J., \& Prakash, M. 2001, Astrophys.J., 550, 426

\bibitem[{Lattimer \& Prakash(2007)}]{NSOEOS}
Lattimer, J.~M., \& Prakash, M. 2007, Phys.Rept., 442, 109

\bibitem[{Levin(1999)}]{LEVIN}
Levin, Y. 1999, The Astrophysical Journal, 517, 328

\bibitem[{Li \& White(2008)}]{MLG}
Li, Y.-S., \& White, S.~D. 2008, Mon.Not.Roy.Astron.Soc., 384, 1459

\bibitem[{Lindblom {et~al.}(1998)Lindblom, Owen, \& Morsink}]{GRINSTHYNS}
Lindblom, L., Owen, B.~J., \& Morsink, S.~M. 1998, Phys. Rev. Lett., 80, 4843

\bibitem[{Lindblom {et~al.}(2000)Lindblom, Owen, \& Ushomirsky}]{SCSRM}
Lindblom, L., Owen, B.~J., \& Ushomirsky, G. 2000, Phys.Rev., D62, 084030

\bibitem[{Lindblom {et~al.}(2001)Lindblom, Tohline, \&
  Vallisneri}]{NOLINEVOLRMODES}
Lindblom, L., Tohline, J.~E., \& Vallisneri, M. 2001, Phys.Rev.Lett., 86, 1152

\bibitem[{Lindblom {et~al.}(2002)Lindblom, Tohline, \&
  Vallisneri}]{NUMNONLINEVOL}
---. 2002, Phys.Rev., D65, 084039

\bibitem[{Mannucci {et~al.}(2008)Mannucci, Maoz, Sharon, Botticella,
  Della~Valle, {et~al.}}]{SNRATE}
Mannucci, F., Maoz, D., Sharon, K., {et~al.} 2008, Mon.Not.Roy.Astron.Soc.,
  383, 1121

\bibitem[{Ott {et~al.}(2006)Ott, Burrows, Thompson, Livne, \& Walder}]{SPINOTT}
Ott, C.~D., Burrows, A., Thompson, T.~A., Livne, E., \& Walder, R. 2006,
  Astrophys.J.Suppl., 164, 130

\bibitem[{Owen(2010)}]{OWENBB}
Owen, B.~J. 2010, Phys.Rev., D82, 104002

\bibitem[{Owen {et~al.}(1998)Owen, Lindblom, Cutler, Schutz, Vecchio,
  {et~al.}}]{GWHYNS}
Owen, B.~J., Lindblom, L., Cutler, C., {et~al.} 1998, Phys.Rev., D58, 084020

\bibitem[{Owen \& Sathyaprakash(1999)}]{MF}
Owen, B.~J., \& Sathyaprakash, B. 1999, Phys.Rev., D60, 022002

\bibitem[{Papaloizou \& Pringle(1978)}]{FIRSTRMODE}
Papaloizou, J., \& Pringle, J. 1978, Mon.Not.Roy.Astron.Soc., 182, 423

\bibitem[{Porquet \& Dubau(1999)}]{GWRMRRS}
Porquet, D., \& Dubau, J. 1999, arXiv:astro-ph/9912065

\bibitem[{Rezania \& Jahan-Miri(2000)}]{VAHID}
Rezania, V., \& Jahan-Miri, M. 2000, Mon.Not.Roy.Astron.Soc., 315, 263

\bibitem[{{Saio}(1982)}]{SAIO}
{Saio}, H. 1982, \apj, 256, 717

\bibitem[{Schenk {et~al.}(2002)Schenk, Arras, Flanagan, Teukolsky, \&
  Wasserman}]{NONLINCOUPLROTNS}
Schenk, A.~K., Arras, P., Flanagan, E.~E., Teukolsky, S.~A., \& Wasserman, I.
  2002, Phys.Rev., D65, 024001

\bibitem[{{Schmidt} {et~al.}(1994){Schmidt}, {Kirshner}, {Eastman}, {Phillips},
  {Suntzeff}, {Hamuy}, {Maza}, \& {Aviles}}]{TYPEIIDIST}
{Schmidt}, B.~P., {Kirshner}, R.~P., {Eastman}, R.~G., {et~al.} 1994, \apj,
  432, 42

\bibitem[{Shibata \& Uryu(2000)}]{SHIBATA}
Shibata, M., \& Uryu, K. 2000, Phys.Rev., D61, 064001

\bibitem[{Spruit \& Phinney(1998)}]{SprPhin}
Spruit, H., \& Phinney, E. 1998, Nature, 393, 139

\bibitem[{Staff {et~al.}(2012)Staff, Jaikumar, Chan, \& Ouyed}]{HKTOP}
Staff, J.~E., Jaikumar, P., Chan, V., \& Ouyed, R. 2012, The Astrophysical
  Journal, 751, 24

\bibitem[{Stergioulas \& Font(2001)}]{NONRELRMODES}
Stergioulas, N., \& Font, J.~A. 2001, Phys.Rev.Lett., 86, 1148

\bibitem[{Thrane \& Coughlin(2013)}]{stochtrack}
Thrane, E., \& Coughlin, M. 2013, arXiv:1308.5292

\bibitem[{Thrane {et~al.}(2011)Thrane, Kandhasamy, Ott, Anderson, Christensen,
  {et~al.}}]{STAMPPAPER}
Thrane, E., Kandhasamy, S., Ott, C.~D., {et~al.} 2011, Phys.Rev., D83, 083004

\bibitem[{{Wagoner}(1984)}]{WAGONER}
{Wagoner}, R.~V. 1984, \apj, 278, 345

\bibitem[{Yakovlev {et~al.}(1999)Yakovlev, Levenfish, \& Shibanov}]{NSCS}
Yakovlev, D., Levenfish, K., \& Shibanov, Y. 1999, Phys.Usp., 42, 737

\bibitem[{Yakovlev \& Pethick(2004)}]{NSC}
Yakovlev, D.~G., \& Pethick, C. 2004, Ann.Rev.Astron.Astrophys., 42, 169

\end{thebibliography}
\end{document}